\pdfoutput=1
\documentclass[12pt]{article}

\font\grande=cmr9.5 scaled \magstep4
\font\medio=cmr9.5 scaled \magstep2
\outer\def\beginsection#1\par{\medbreak\bigskip
      \message{#1}\leftline{\bf#1}\nobreak\medskip
\vskip-\parskip
      \noindent}
\usepackage{graphicx} 
\begin{document}
\bibliographystyle {unsrt}

\titlepage

\begin{flushright}
CERN-TH-2018-107
\end{flushright}

\vspace*{1.5cm}
\begin{center}
{\grande The propagating speed of relic gravitational waves}\\
\vspace{6mm}
{\grande and their refractive index during inflation}\\
\vspace{15mm}
 Massimo Giovannini 
 \footnote{Electronic address: massimo.giovannini@cern.ch} \\
\vspace{1cm}
{{\sl Department of Physics, 
Theory Division, CERN, 1211 Geneva 23, Switzerland }}\\
\vspace{0.5cm}
{{\sl INFN, Section of Milan-Bicocca, 20126 Milan, Italy}}

\vspace*{1cm}
\end{center}

\centerline{\medio  Abstract}
\vspace{5mm}
If the refractive index of the tensor modes 
increases during a conventional inflationary stage of expansion
the relic graviton spectrum is tilted towards high frequencies. 
Two apparently diverse parametrizations of this effect are 
shown to be related by a rescaling of the four-dimensional 
metric through a conformal factor that involves the refractive index itself. 
Non-monotonic spectra with a maximum in the MHz region correspond 
to a limited variation of the refractive index terminating well before the end of inflation. 
After exploring a general approach encompassing the ones proposed so far, 
we estimate the required sensitivity for the direct detection of the 
predicted gravitational radiation and demonstrate that the allowed regions 
of the parameter space are within reach for some of the 
planned detectors operating either in the audio band (like Ligo/Virgo 
and Kagra) or in the mHz band (like Lisa, Bbo and Decigo).

\vskip 0.5cm

\nonumber
\noindent

\vspace{5mm}

\vfill
\newpage

\renewcommand{\theequation}{1.\arabic{equation}}
\setcounter{equation}{0}
\section{Introduction}
\label{sec1}
The tensor modes of the geometry are known to be parametrically amplified in the early Universe thanks 
to the pumping action of the gravitational field,  as suggested by Grishchuk well before the 
formulation of the conventional inflationary paradigm \cite{zeroa,zerob}. 
According to the quantum theory of parametric amplification (originally developed in the case of 
optical photons \cite{glauber}), relic gravitons are produced from the vacuum  or from a specific initial state with opposite 
momenta \cite{zeroc}. Squeezed states of relic gravitons are characteristic of many scenarios and, in particular, 
of conventional inflationary models \cite{zerod,zeroe}. The degree of quantum coherence of the large-scale correlations 
could be used to disambiguate their origin, at least in principle \cite{zerof}.

Gravitational waves might acquire an effective index of refraction when they travel in curved 
space-times \cite{onea,oneb} and this possibility has been recently revisited by 
studying the parametric amplification of the tensor modes 
of the geometry during a quasi-de Sitter stage of expansion \cite{two}:
when the refractive index mildly increases during inflation  
the corresponding speed of propagation of the waves diminishes and 
the power spectra of the relic gravitons are then
blue, i.e. tilted towards high frequencies. After the analysis of \cite{two} the same idea 
has been pursued in two similar papers \cite{three}. The first 
goal of the present analysis is a specific discussion of the relation between the approaches 
of Refs. \cite{two} and \cite{three} which ultimately coincide since they are related by a
 conformal rescaling involving the refractive index. We shall then explore a more 
general approach encompassing the ones discussed previously and analyze 
the actual phenomenological constraints together with the associated detectability 
prospects.

A refractive index varying in the early Universe is implicitly contained in the 
work of Szekeres \cite{onea,oneb} aiming at the general formulation of a macroscopic theory of gravitation. 
An effective index of refraction  may also arise in modified theories of gravity violating the equivalence principle \cite{two,three}.
In the model-independent perspective pursued in Ref. \cite{two} and subsequently followed in \cite{three} (see, in particular, the second paper) the pivotal parameter of the discussion will therefore be the rate of variation 
of the refractive index in units of the Hubble rate, conventionally denoted by $\alpha$ and defined by: 
\begin{equation}
\alpha = \frac{n \, H}{\dot{n}}, \qquad \epsilon = - \frac{\dot{H}}{H^2}\ll 1,\qquad c_{gw}(t) = \frac{c}{n(t)}.
\label{ref1}
\end{equation}
In Eq. (\ref{ref1}) the overdot denotes a derivation\footnote{In section 
\ref{sec3} the overdot will be used to denote a derivation with respect to the $\eta$-time which is related to the conformal time 
coordinate as $n(\eta) d\eta = d\tau$ where $n(\eta)$ is the refractive index.} 
with respect to the cosmic time coordinate $t$; $\epsilon$ and $c_{gw}(t)$ are, respectively, the standard slow-roll parameter $\epsilon$ and the 
phase velocity of the relic gravitons in units of the speed of light.  The phase velocity coincides, in the present case,  
with the group velocity and if the refractive index increases during inflation 
neither the phase velocity nor the group velocity will be superluminal. Note finally that the conventional 
slow-roll dynamics does not necessarily imply the validity of the so-called consistency relations which 
are instead broken by the presence of the refractive index so that the tensor spectral index and the 
tensor-to-scalar ratio are not solely determined by $\epsilon$, as it happens when the consistency 
relations are not violated.

During the last year the Ligo/Virgo collaboration published the three-detector observations 
of gravitational waves from black-hole coalescence \cite{LIGO1}, the evidence of gravitational waves 
from neutron star inspiral \cite{LIGO2}, and the observation of a $50$-solar-mass binary black hole 
coalescence for a redshift $z =0.2$ \cite{LIGO3}. As gravitational wave astronomy is opening a new 
window for the observation of the present Universe, the theory is beginning to consider 
more concretely the potential implications of the current developments for the early Universe.
In the course of inflation, the evolution of the space-time curvature induces 
a stochastic background of relic gravitons \cite{onea,oneb} with a spectral energy 
density extending today from frequencies ${\mathcal O}(\mathrm{aHz})$ (i.e. $1 \,\mathrm{aHz} = 
10^{-18}\, \mathrm{Hz}$) up to frequencies ${\mathcal O}(\mathrm{GHz})$ (i.e. $1\, \mathrm{GHz} = 10^{9} \mathrm{Hz}$).  
In the conventional scenario the absolute normalization of the tensor power spectrum solely depends on the tensor to scalar ratio
 Here we shall be assuming the limit $r_{T}<0.07$ for the tensor spectral index 
 from a joint analysis of Planck and BICEP2/Keck array data~\cite{BICPL} (see also \cite{first}). The typical energy density of inflationary gravitational waves between the mHz  (i.e. $1 \,\mathrm{mHz} = 10^{-3}$ Hz) and the kHz is, optimistically, ${\mathcal O}(10^{-16.5})$ in terms of $\Omega_{\mathrm{gw}}(\nu,\tau_{0})$ 
which will denote throughout the spectral energy density in critical units at the present (conformal) time $\tau_{0}$.
The corresponding chirp amplitude $h_{c}(\nu,\tau_{0})$
will therefore be ${\mathcal O}(10^{-29})$ for a typical comoving frequency of $0.1$ kHz as it follows 
from the general relation:
\begin{equation}
h_{c}(\nu,\tau_{0}) = 1.263 \times 10^{-20} \biggl(\frac{100 \,\, \mathrm{Hz}}{\nu}\biggr) \, \sqrt{h_{0}^2 \,\Omega_{\mathrm{gw}}(\nu,\tau_{0})},
\label{B}
\label{ref2}
\end{equation} 
connecting the chirp amplitude to the spectral energy density in critical units.
While the $h_{c}$ corresponding to the astronomical signals detected by the Ligo/Virgo collaboration is ${\mathcal O}(10^{-21})$ 
\cite{LIGO1,LIGO2,LIGO3}, various improvements are expected at least in the so-called 
audio band between few Hz and $10$ kHz. In this perspective it is useful to discuss potential deviations from the 
conventional inflationary signal hoping for larger spectral amplitudes over intermediate and high frequencies. 

The plan of this paper is the following. In section \ref{sec2} we shall demonstrate that 
the apparently different actions proposed so far to account for the same effects 
are indeed equivalent up to conformal rescaling 
of the four-dimensional metric. A general approach to the analysis of the power spectrum
is proposed in section \ref{sec3} where the tensor to scalar ratio and the relevant power spectra are 
explicitly derived. In section \ref{sec4} 
the phenomenological constraints relevant to the present scenario will be  studied by charting the 
corresponding parameter space; the potential signals will be
confronted with the hoped sensitivities of the wide-band interferometers on the ground and in space.  
The concluding remarks are collected in section \ref{sec5} and some useful details are
relegated to the appendix \ref{APPA} for the interested readers. 

\newpage
\renewcommand{\theequation}{2.\arabic{equation}}
\setcounter{equation}{0}
\section{Actions and their parametrization}
\label{sec2}
Let us start with an obvious remark and recall that, in the Einstein frame, the metric can be separated into a background value supplemented by the inhomogeneous contribution:
\begin{equation}
g_{\mu\nu}^{(E)} = \overline{g}_{\mu\nu}^{(E)} + \delta_{\mathrm{t}} g_{\mu\nu}^{(E)},
\label{EIN1}
\end{equation}
where  $\delta_{t} g_{\mu\nu}^{(E)}$ denotes the tensor fluctuations of the geometry; 
in the present context the vector modes are anyway
suppressed and shall be completely disregarded while the scalar fluctuations are described in terms 
of the gauge-invariant curvature inhomogeneities (see e.g. Eqs. (\ref{ST1C})--(\ref{ST1Ca}) 
and discussion therein) but will not play a direct role in the forthcoming considerations since the refractive 
index is fully homogeneous. 

Even if the analysis of the present section does not demand the absence of the intrinsic curvature, 
we shall now restrict the attention to the case where $\overline{g}_{\mu\nu}^{(E)}$ is conformally flat i.e.
\begin{equation}
\overline{g}_{\mu\nu}^{(E)}= a^2_{E}(\tau) \eta_{\mu\nu}, \qquad \delta_{\mathrm{t}} g_{ij} = - a^2_{E}(\tau) h_{ij}^{(E)},
\label{EIN2}
\end{equation}
where $\eta_{\mu\nu}=\mathrm{diag}(+1, \, -1,\, -1,\, -1)$ is the Minkowski metric 
with signature mostly minus and the three-dimensional rank-two tensor $h_{ij}^{(E)}$ is both divergenceless and traceless.
In Eq. (\ref{EIN2}) and throughout the rest of the paper $\tau$ denotes the conformal time 
coordinate. In the frame defined by Eq. (\ref{EIN1}) the action for the tensor modes of the geometry has been 
derived long ago by Ford and Parker in Ref. \cite{four,four2}:
\begin{equation}
S^{(E)} = \frac{1}{8 \ell_{P}^2} \int d^{4} x \, \sqrt{ - \overline{g}_{(E)}} \, \overline{g}^{\,\mu\nu}_{(E)} \partial_{\mu} 
h^{(E)}_{ij} \partial_{\nu} h^{(E)}_{ij}, \qquad \ell_{P} = \sqrt{8 \pi G} = \frac{1}{\overline{M}_{P}}.
\label{EIN3}
\end{equation}
Equation (\ref{EIN3}) can be supplemented by a second term containing the dependence 
on the refractive index $n(\tau)$ which is the inverse of the phase velocity\footnote{In the present case the phase velocity
coincides with the  group velocity since the relation between frequencies and wavenumbers is linear (as in the case of Refs.  \cite{two,three}) but the refractive 
index is time-dependent. } in natural units $\hbar= c =1$, i.e. $c_{gw}(\tau) = 1/n(\tau)$:
\begin{equation}
S^{(E)} = \frac{1}{8 \ell_{P}^2} \int d^{4} x \sqrt{-\overline{g}_{(E)}} \biggl\{ \overline{g}^{\mu\nu}_{(E)} \partial_{\mu} h^{(E)}_{ij} \partial_{\nu} h^{(E)}_{ij}
+ [c_{gw}^2(\tau) -1 ]\overline{P}^{\mu\nu}_{(E)} \partial_{\mu} h^{(E)}_{ij} \partial_{\nu} h^{(E)}_{ij} \biggr\},
\label{EIN4}
\end{equation}
where  $\overline{P}^{\mu\nu}_{(E)}$ is the conventional spatial projector tensor orthogonal to $\overline{u}^{(E)}_{\mu}$: 
\begin{equation}
\overline{P}^{\mu\nu}_{(E)} = \overline{g}^{\mu\nu}_{(E)} - \overline{u}^{\mu}_{(E)} \overline{u}^{\nu}_{(E)}, \qquad 
\overline{g}^{\mu\nu}_{(E)} \overline{u}^{(E)}_{\mu} \overline{u}^{(E)}_{\nu} =1.
\label{EIN5}
\end{equation}
Since in comoving coordinates $\overline{u}^{(E)}_{\mu} = \overline{g}^{(E)}_{\mu0}/\sqrt{\overline{g}^{(E)}_{00}}$ and 
$\overline{u}^{\mu}_{(E)}= \delta^{\mu}_{0}/\sqrt{\overline{g}^{(E)}_{00}}$ the action of Eq. (\ref{EIN4}) after straightforward modifications assumes the following form \cite{two}:
\begin{equation}
S^{(E)} =  \frac{1}{8 \ell_{P}^2} \int d^{3} x \int d\tau \,\,a^2_{E} \,\,\biggl[ \partial_{\tau} h^{(E)}_{ij} \partial_{\tau} h^{(E)}_{ij} - 
c_{gw}^2(\tau) \partial_{k} h^{(E)}_{ij} \partial^{k} h^{(E)}_{ij} \biggr].
\label{EIN6}
\end{equation}
When the refractive index grows in time \cite{two} $c_{gw}(\tau)=1/n(\tau)$ decreases but what matters is that thanks either to a conformal rescaling or to a field 
redefinition the action (\ref{EIN6}) can be brought into the following generalized form: 
\begin{equation}
S =  \frac{1}{8 \ell_{P}^2} \int d^{3} x \int d\tau \biggl[ A(\tau) \partial_{\tau} h_{ij} \partial_{\tau} h_{ij}  - B(\tau) \partial_{k} h_{ij} \partial^{k} h_{ij}  + C(\tau) h_{ij} h_{ij} \biggr],
\label{EIN7}
\end{equation}
where $A(\tau)$, $B(\tau)$ and $C(\tau)$ depend on the conformal time coordinate $\tau$ and are themselves functionals of either 
the scale factor, or of the refractive index or even of both \cite{two}. A term like $C(\tau)$ may appear by a banal field 
redefinition of the type $a_{E} h^{(E)}_{ij} = h_{ij}$ so we shall not dwell about this issue and focus hereunder on $A(\tau)$ and $B(\tau)$.
After the results of Ref. \cite{two}, the authors of Ref. \cite{three} suggested the following form of the action:
\begin{equation}
S = \frac{1}{8 \ell_{P}^2}  \int d^{3} x \int d\tau a^2 \biggl[ n^2(\tau) \partial_{\tau} h_{ij} \partial_{\tau} h_{ij} - \partial_{k} h_{ij} \partial_{k} h_{ij} \biggr],
\label{EIN8}
\end{equation}
and argued that it leads to a blue spectrum of tensor modes\footnote{More precisely, while the first paper of Ref. \cite{three}
did not even mention the analog results of Ref. \cite{two}, the second paper of Ref. \cite{three} did allude to Ref. \cite{two} 
by claiming (in a footnote) that the  two approaches are radically different. In what follows 
we shall instead demonstrate that Eqs. (\ref{EIN4})--(\ref{EIN6}) and (\ref{EIN7})--(\ref{EIN8}) are 
exactly the same up to a conformal rescaling.}. It is relevant to point out that 
Eq. (\ref{EIN6}) has been used in Ref. \cite{two} to argue that the spectra of relic gravitons may 
indeed increase in the presence of an inflating background. The authors of Ref. \cite{three} 
did consider instead Eq. (\ref{EIN8}), reached a similar conclusion, but did not realize (or acknowledge) that Eq. (\ref{EIN8}) 
is conformally related to Eq. (\ref{EIN6}). As a consequence  Ref. \cite{three} analyzed the case 
where the background is inflating but in the frame defined by Eq. (\ref{EIN8}). 
The choice of Ref. \cite{three} is not particularly consistent and it is even confusing 
in the light of the previous analyses of Ref. \cite{two}. We shall propose a generalized parametrization 
which may hopefully address and clarify the potentially confusing statements present in the 
current literature (see, in particular, Eq. (\ref{fifth}) and discussion therein).

To show the conformal equivalence of Eqs. (\ref{EIN6}) and (\ref{EIN8}) let us now define a frame conformally related to  Eq. (\ref{EIN1}) 
by positing 
\begin{equation}
g_{\mu\nu}^{(E)} = \Omega^2(\tau) G_{\mu\nu}, \qquad \sqrt{- g_{(E)}} = \Omega^4(\tau) \, \sqrt{-G}.
\label{G1}
\end{equation}
Exactly as in the case of Eq. (\ref{EIN1}), the  $G_{\mu\nu}$ can be decomposed into a homogeneous part 
supplemented by its own tensor inhomogeneities: 
\begin{equation}
G_{\mu\nu} = \overline{G}_{\mu\nu}+ \delta_{t} G_{\mu\nu},\qquad \overline{G}_{\mu\nu}= a^2(\tau) \eta_{\mu\nu}.
\label{G2}
\end{equation}
In the frame defined by Eqs. (\ref{G1}) and (\ref{G2})  the tensor modes of the geometry obey: 
\begin{equation}
\delta_{\mathrm{t}} G_{ij} = - a^2(\tau) h_{ij}, \qquad \partial_{i} h^{i}_{j} = h_{i}^{i}.
\label{G3}
\end{equation}
The tensor modes of Eq. (\ref{G3}) coincide with $h_{ij}^{(E)}$, i.e. $h_{ij}^{(E)} = h_{ij}$:
this means that the tensor fluctuations are not only invariant under infinitesimal coordinate 
transformations but they are equally invariant under conformal rescaling. 
To appreciate this well known aspect we can take the tensor fluctuations of both sides of Eq. (\ref{G2}) and obtain:
\begin{equation}
\delta_{\mathrm{t}} g_{\mu\nu}^{(E)} = \Omega^2 \delta_{t} G_{\mu\nu}, \qquad \overline{g}^{(E)}_{\mu\nu} = \Omega^2(\tau) \overline{G}_{\mu\nu}. 
\label{G4}
\end{equation}
Inserting Eqs. (\ref{EIN2}) and (\ref{G2}) into Eq. (\ref{G4}) we simply have, as anticipated, that 
\begin{equation}
h_{ij}^{(E)} = h_{ij}, \qquad a_{E}(\tau) = \Omega(\tau) a(\tau).
\label{G5}
\end{equation}
Equation (\ref{G5}) implies that if the two backgrounds are conformally related the gauge-invariant tensor 
amplitudes are also the same in the two frames. A similar property holds for the curvature perturbations 
on comoving orthogonal hypersurfaces\footnote{The frame-invariance of the scalar modes 
is less immediate than for the tensors. However it can be shown that the gauge-invariant curvature inhomogeneities (i.e. 
the curvature perturbations on comoving orthogonal hypersurfaces) are frame invariant (see in this respect the discussion 
in Eqs. (\ref{SSS1}), (\ref{SSS2}) and (\ref{SSS5})).}. Bearing in mind the results of Eqs. (\ref{G4}) and (\ref{G5}) 
we can perform the conformal rescaling of the projector and of the four-velocities:
\begin{equation}
\overline{P}^{\mu\nu}_{(E)} = \frac{1}{\Omega^2} \overline{P}^{\mu\nu}, \qquad \overline{P}^{\mu\nu} = \overline{G}^{\mu\nu} - \overline{U}^{\mu} \overline{U}^{\nu}, \qquad \overline{G}_{\mu\nu} \overline{U}^{\mu} \overline{U}^{\nu} =1,\qquad  \overline{U}^{\mu} = \frac{ \overline{u}^{\mu}_{(E)}}{\Omega},
\label{G6}
\end{equation}
so that, thanks to Eqs. (\ref{G4}), (\ref{G5}) and (\ref{G6}) the action (\ref{EIN1}) in the conformally related frame becomes:
\begin{equation}
S^{(E)} \to S= \frac{1}{8 \ell_{P}^2} \int d^{4} x \biggl\{ \sqrt{-\overline{G}} \biggl[ \overline{G}^{\mu\nu} \Omega^2 \partial_{\mu} h_{ij} \partial_{\nu} h_{ij} + [c_{gw}^2(\tau) -1] \Omega^2 \overline{P}^{\mu\nu}\partial_{\mu} h_{ij} \partial_{\nu} h_{ij} \biggr]\biggr\}.
\label{G7}
\end{equation}
If we now identify the conformal factor with the refractive index itself 
\begin{equation}
\Omega(\tau)  = \frac{1}{c_{gw}(\tau)} = n(\tau), 
\label{G7a}
\end{equation}
the action of Eq. (\ref{G7}) becomes 
\begin{equation}
 S = \frac{1}{8 \ell_{P}^2}  \int d^{3} x \int d\tau a^2 \biggl[ n^2(\tau) \partial_{\tau} h_{ij} \partial_{\tau} h_{ij} - \partial_{k} h_{ij} \partial_{k} h_{ij} \biggr],
\label{G8}
\end{equation}
which is exactly the action anticipated in Eq. (\ref{EIN8}) and discussed in Ref. \cite{three}. All in all the action of Eqs. (\ref{EIN4}) 
and (\ref{EIN6}) (originally discussed in \cite{two}) and the action (\ref{G8}) (subsequently used in Ref. \cite{three}) are one and the same 
action since they are simply related by a conformal rescaling.

To make even more transparent the equivalence of the two descriptions it is useful to write down explicitly 
the evolution of the tensor modes in the two frames. From Eq. (\ref{EIN8}) it follows 
that the evolution of $h_{ij}^{(E)}$ is given by: 
\begin{equation}
h_{ij}^{(E)\, \prime\prime} + 2 \frac{a_{E}^{\prime}}{a_{E}} h_{ij}^{(E)\, \prime} - \frac{\nabla^2 h_{ij}^{(E)}}{n^2(\tau)} =0.
\label{G9}
\end{equation}
Similarly, the equation for $h_{ij}$ can be deduced from Eq. (\ref{G8}) and the result will be:
\begin{equation}
h_{ij}^{\prime\prime} + 2 \biggl(\frac{a^{\prime}}{a} + \frac{n^{\prime}}{n} \biggr) h_{ij}^{ \prime} - \frac{\nabla^2 h_{ij}}{n^2(\tau)} =0.
\label{G10}
\end{equation}
By virtue of Eq. (\ref{G5}) we see that Eqs. (\ref{G9}) and (\ref{G10}) are two ways of writing the same equation 
since 
\begin{equation}
\frac{a_{E}^{\prime}}{a_{E}} = \frac{a^{\prime}}{a} + \frac{n^{\prime}}{n}, \qquad h_{ij}^{(E)} = h_{ij}.
\label{G11}
\end{equation} 
We must finally remark that $n(\tau)$ and $c_{gw}(\tau)$ are completely homogeneous. 
While this is the simplest situation, if the spectral index is not fully homogeneous there will be additional 
scalar fluctuations mixing with the scalar adiabatic mode. At 
the end of appendix \ref{APPA} it is demonstrated that the curvature perturbations share 
the same property of the tensor amplitudes since they are gauge-invariant and also invariant under 
conformal rescaling.

\renewcommand{\theequation}{3.\arabic{equation}}
\setcounter{equation}{0}
\section{Generalized form of the tensor to scalar ratio}
\label{sec3}
\subsection{Diagonal form of the action}
If we conventionally define the pivotal frame as the one where the background experiences a phase a quasi-de Sitter expansion,
we should conclude that the choice of the pivotal frame is to some extent ambiguous, as already suggested. This ambiguity 
is not of a fundamental nature but simply stems from the current literature. Indeed, as already mentioned, in Ref. \cite{two} 
the action (\ref{EIN6}) has been used to argue that the dynamics of the refractive index during 
an inflationary stage induces a growing spectrum of the tensor modes. After Ref. \cite{two} the authors 
of Ref. \cite{three} suggested instead the action (\ref{EIN8}) and considered (in their own framework)
an inflating background. The authors of Ref. \cite{three} did not acknowledge the results of Ref. \cite{two} 
and did not notice or appreciate that the actions (\ref{EIN6}) and (\ref{EIN8}) are one and the same action, up to a conformal rescaling: this 
is the main reason of the frame ambiguity mentioned above. 
It is not a surprise that if we impose that the background in inflating in both frames defined 
by Eqs.  (\ref{EIN6}) and (\ref{EIN8}), growing spectra of the tensor modes will be obtained in both frames.
In what follows the aim will be first to quantize the problem in general terms and then 
to deduce the spectral energy density in various cases that have not been addressed so far.
For this purpose and to account of the perspectives of Refs. \cite{two,three}
it is convenient to analyze simultaneously the two situations by considering the following diagonal form of the action for the tensor 
modes\footnote{For the present discussion the relevant phenomenological case is the one where the background geometry 
undergoes a conventional inflationary expansion, even if the discussion of this section also applies to different 
situations.}:
\begin{equation}
S= \frac{1}{8 \ell_{P}^2} \int d^{4} x\, a^2 n^{2 \gamma} \biggl[ \partial_{\tau} h_{ij} \partial_{\tau} h_{ij}  -  \frac{1}{n^2}
\partial_{k} h_{ij} \partial_{k} h_{ij}\biggr],
\label{fifth}
\end{equation}
where $\gamma$ is a further parameter varying between $0$ and $1$.
If $\gamma=0$ the action (\ref{fifth}) reproduces exactly Eq. (\ref{EIN6}) with 
$c_{gw}(\tau) = 1/n(\tau)$. Conversely when $\gamma = 1$ Eq. (\ref{fifth}) 
reproduces Eq. (\ref{G8}) (or Eq. (\ref{EIN8})). According to a purely heuristic logic the values of $\gamma$ may also lie outside 
the interval $0\leq \gamma \leq 1$: in this perspective $\gamma$ can be considered as a further 
parameter that simply rescales the tensor spectral index from its value for $\gamma =0$. However, if 
$\gamma \neq 0$ the sign of the tensor spectral index does not change, as we shall see.
The presence of the refractive index in the second term inside the squared bracket  
suggests that the action can be diagonalized  in terms of a new time coordinate, 
namely $d\tau = n(\eta) d\eta$. In the $\eta$-parametrization  the action of Eq. (\ref{fifth}) 
becomes:
\begin{equation}
S = \frac{1}{8 \ell_{P}^2} \int d^{3}x \, d\eta \,\, b^2(\eta)  \biggl[ \partial_{\eta} h_{ij} \partial_{\eta} h_{ij}  -  \partial_{k} h_{ij} \partial_{k} h_{ij}\biggr], \qquad 
b(\eta) = a(\eta) \, n^{\gamma -1/2}(\eta).
\label{sixth}
\end{equation}
While the relation of $b(\eta)$ to $a(\eta)$ and $n(\eta)$ may change, what matters is the overall dependence of the power spectrum on the background fields. The parametrization of Eq. (\ref{sixth}) will be used to demonstrate, in a unified approach to the problem, that both Eqs. (\ref{EIN8}) and (\ref{G8}) lead
to a growing power spectrum for typical wavelengths larger than the Hubble radius. 

\subsection{Quantization and power spectrum}
For a correct normalization of the power spectrum the quantization of the system is essential 
and for this purpose the $\eta$-parametization is more convenient, as already emphasized in 
Ref. \cite{two}  without specific details. This gap will now be bridged by only 
using the action of Eq. (\ref{sixth}) for the explicit derivation of the power spectrum, of the tensor to scalar ratio and of 
all the other relevant quantities. We start by noticing that 
from the explicit form of the tensor polarizations given in Eq. (\ref{ST0}) 
it is convenient to introduce $h_{\lambda}(\vec{x},\eta)$
\begin{equation}
h_{ij}(\vec{x},\eta) = \sqrt{2} \,\ell_{P}\sum_{\lambda = \oplus, \, \otimes} h_{\lambda}(\vec{x},\eta) \, e^{(\lambda)}_{ij}.
\label{Q1}
\end{equation}
Inserting Eq. (\ref{Q1}) into Eq. (\ref{sixth}) the action becomes 
\begin{equation}
S = \frac{1}{2} \sum_{\lambda} \int d^{3}x\, d\eta \,\, b^2(\eta)\,\,\biggl[ \dot{h}_{\lambda}^2 - \partial_{k} h_{\lambda} \partial^{k} h_{\lambda} \biggr],
\label{Q2}
\end{equation}
where the overdot denotes a derivation with respect to $\eta$, i.e. $\dot{h}_{\lambda} = \partial_{\eta} h_{\lambda}$.
From the definition of the canonical momenta, the Hamiltonian in the $\eta$-parametrization becomes\footnote{An $\eta$-dependent canonical transformations will change the form of the Hamiltonian (\ref{Q3}); the resulting Hamiltonian will then follow 
from the same action sixth where, however, some total derivative terms have been subtracted or added. }: 
\begin{equation}
H(\eta) = \frac{1}{2}  \sum_{\lambda} \int d^{3}x \, \biggl[ \frac{\pi_{\lambda}^2}{b^2(\eta)} + b^2(\eta) \partial_{k} h_{\lambda} \partial^{k} h_{\lambda} \biggr], \qquad \pi_{\lambda} = b^2(\eta) \dot{h}_{\lambda}.
\label{Q3}
\end{equation}
The canonical form of the commutation relations at equal $\eta$-times
\begin{equation}
[\hat{h}_{\lambda}(\vec{x}, \eta), \, \hat{\pi}_{\lambda^{\prime}}(\vec{y}, \eta) ] = i\, \delta^{(3)}(\vec{x} - \vec{y}) \, \delta_{\lambda\, \lambda^{\prime}},
\label{Q5a}
\end{equation}
and the explicit form of the Hamiltonian (\ref{Q3}) lead  directly to the evolution equations for $\hat{h}_{\lambda}$ and $\hat{\pi}_{\lambda}$:
\begin{equation}
\partial_{\eta} \hat{h} = i [\hat{H}, \hat{h}_{\lambda}] = \frac{\hat{\pi}_{\lambda}}{b^2}, \qquad
\partial_{\eta} \hat{\pi} = i [\hat{H}, \hat{\pi}_{\lambda}] = b^2 \nabla^2 \hat{h}_{\lambda}.
\label{Q5}
\end{equation}
Using Eq. (\ref{Q5}) the Fourier representations of the operators $\hat{h}_{\lambda}$ and $\hat{\pi}_{\lambda}$ is:
\begin{eqnarray}
\hat{h}_{\lambda}(\vec{x},\eta) = \frac{1}{(2\pi)^{3/2}} \int d^{3} k \biggl[ \hat{a}_{\vec{k},\, \lambda} \, F_{k,\, \lambda} e^{- i \vec{k} \cdot \vec{x}} +  \hat{a}^{\dagger}_{\vec{k},\, \lambda} \, F^{*}_{k,\, \lambda} e^{- i \vec{k} \cdot \vec{x}} \biggr],
\label{Q6}\\
\hat{\pi}_{\lambda}(\vec{x},\eta) = \frac{1}{(2\pi)^{3/2}} \int d^{3} k \biggl[ \hat{a}_{\vec{k},\, \lambda} \, G_{k,\, \lambda} e^{- i \vec{k} \cdot \vec{x}} +  \hat{a}^{\dagger}_{\vec{k},\, \lambda} \, G^{*}_{k,\, \lambda} e^{- i \vec{k} \cdot \vec{x}} \biggr],
\label{Q7}
\end{eqnarray}
where $[ a_{{\vec{k}, \, \lambda}}, \, \hat{a}^{\dagger}_{{\vec{p}, \, \lambda}}] = \delta^{(3)}(\vec{k}-\vec{p}) \, \delta_{\lambda,\, \lambda^{\prime}}$; 
the evolution of the mode functions $F_{k,\lambda}$ and $G_{k,\lambda}$  and the normalization of their Wronskian 
immediately follows from Eqs. (\ref{Q5a})--(\ref{Q5}) and (\ref{Q6})--(\ref{Q7}):
\begin{eqnarray}
&& \dot{F}_{k,\,\lambda} = \frac{G_{k,\,\lambda}}{b^2}, \qquad \dot{G}_{k,\,\lambda} = - k^2 b^2 F_{k,\,\lambda}
\label{Q8}\\
&& F_{k,\, \lambda}(\eta) G^{*}_{k,\, \lambda}(\eta) - F_{k,\, \lambda}^{*}(\eta) G_{k,\, \lambda}(\eta) = i.
\label{Q9}
\end{eqnarray}
The evolution equations of the mode functions can be decoupled as:
\begin{equation}
\ddot{F_{k}} + 2 \frac{\dot{b}}{b} \dot{F}_{k} + k^2 F_{k} =0, \qquad G_{k}= b^2 \dot{F}_{k},
\label{Q10}
\end{equation}
where the polarization index has been omitted since the result of Eq. (\ref{Q10}) holds both for $\oplus$ and for $\otimes$.
The quantum mechanical initial conditions correspond to 
\begin{equation}
\lim_{\eta \to - \infty} F_{k}(\eta) \to \frac{1}{\sqrt{2 k } b(\eta)} e^{- i k \eta}.
\label{Q10a}
\end{equation}
In the limit $n(\eta) \to 1$ the $\eta$-time turns into the conformal time while $b(\eta) \to a(\tau)$: therefore 
the initial data of Eq. (\ref{Q10a}) generalize the standard quantum mechanical initial condition to the case where 
the refractive index is dynamical. Note that since $b(\eta)$ has a power-law dependence 
for $\eta < - \eta_{*}$ (see below Eqs. (\ref{MF3a}) and (\ref{MF4})), Eq. (\ref{Q10}) can be 
exactly solved in the $\eta$-time (see Eq. (\ref{MF5})) with the initial conditions 
dictated by Eq. (\ref{Q10a}).

In conclusion we have that the full mode expansion of the tensor amplitude $\hat{h}_{ij}(\vec{x},\eta)$ in the 
$\eta$-time parametrization easily follows from the combination of Eq. (\ref{Q1}) and (\ref{Q6}) and 
it is given by:
\begin{equation}
\hat{h}_{ij}(\vec{x},\eta) = \frac{\sqrt{2} \ell_{P}}{(2\pi)^{3/2}}\sum_{\lambda} \int \, d^{3} k \,\,e^{(\lambda)}_{ij}(\vec{k})\, [ F_{k,\lambda}(\eta) \hat{a}_{\vec{k}\,\lambda } e^{- i \vec{k} \cdot \vec{x}} + F^{*}_{k,\lambda}(\eta) \hat{a}_{\vec{k}\,\lambda }^{\dagger} e^{ i \vec{k} \cdot \vec{x}} ],
\label{Q11}
\end{equation}
The two-point function computed from Eqs. (\ref{Q11}) is simply\footnote{For the sake of notational 
accuracy, we remind that, throughout this analysis, natural logarithms will be denoted by $\ln$ while the common logarithms 
will be denoted by $\log$.}
\begin{equation}
\langle \hat{h}_{ij}(\vec{x}, \eta) \, \hat{h}_{ij}(\vec{x}+ \vec{r},\eta) \rangle = \int d\ln{k} \, {\mathcal P}_{T}(k,\eta) j_{0}(k r), 
\label{Q12a}
\end{equation}
where $j_{0}(kr)$ is the spherical Bessel function of zeroth order \cite{abr1,abr2}. The tensor power spectrum of Eq. (\ref{Q12a}) is 
then given by
\begin{equation}
{\mathcal P}_{T}(k,\eta) = \frac{4 \ell_{P}^2}{\pi^2} k^3 |F_{k}(\eta)|^2.
\label{MF1b}
\end{equation} 
  
\subsection{Explicit evolution of the mode functions and related matters}
\begin{figure}[!ht]
\centering
\includegraphics[height=5cm]{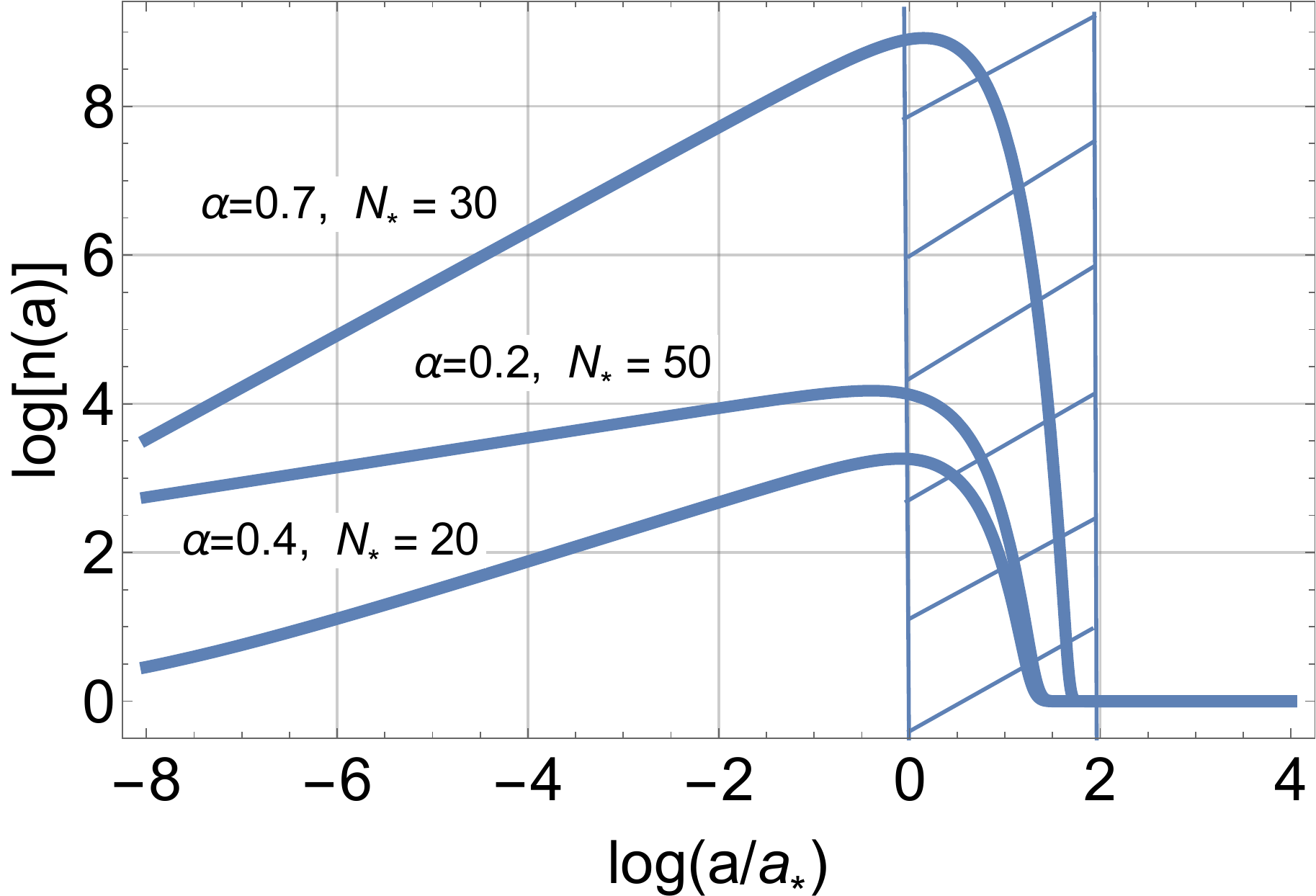}
\includegraphics[height=5cm]{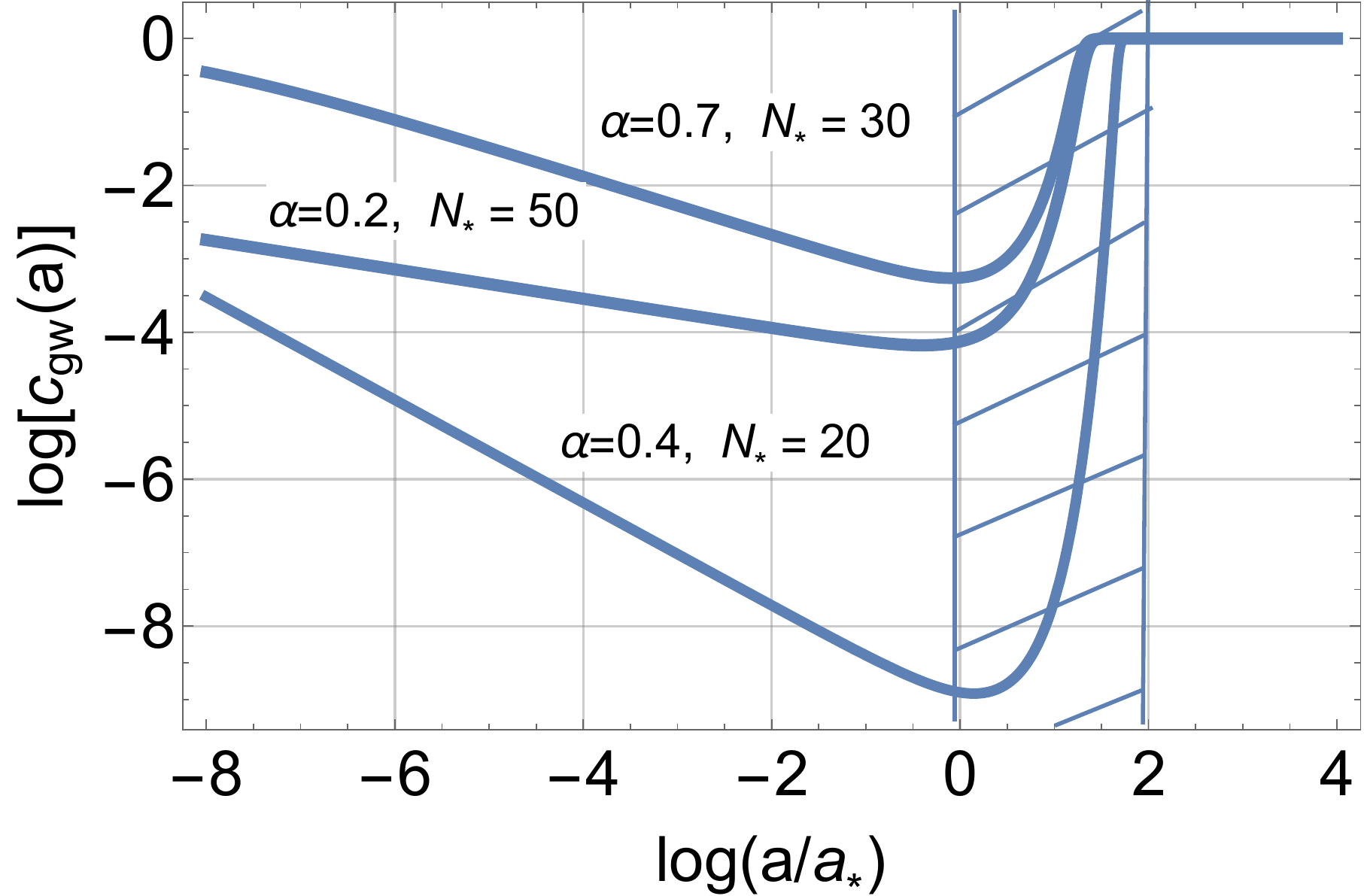}
\caption[a]{In the left plot the common logarithm of the refractive index as a function of the scale factor is illustrated for two 
profiles characterized by different values of $\alpha$ and of $N_{*}$. In the right plot we report instead the common logarithm of the 
propagating speed (coinciding, in our case, with the phase velocity) in natural units.
On the horizontal axes of both plots the common logarithm of the scale factor is reported.}
\label{FIGU1}      
\end{figure}

As already mentioned in Eq. (\ref{ref1}) the variation of the refractive index is measured in units of the Hubble rate 
and for the sake of concreteness we shall consider the general situation where 
\begin{equation}
n(a) = n_{i} \biggl(\frac{a}{a_{i}}\biggr)^{\alpha}, \qquad a< a_{*},
\label{NNA1}
\end{equation}
while $n(a) = 1$ for $a> a_{*}$. It is not difficult to imagine various continuous interpolation between 
the two regimes but what matters is the continuity of $n(a)$, not the specific form
of the profile across the normalcy transition. One of the simplest possibilities is given by\footnote{Note that $n_{i} \geq 1$ but we shall 
always consider the case $n_{i}=1$ as representative of the general situation.}
$n(a,\xi) = n_{i} (a/a_{i})^{\alpha} e^{- \xi a/a_{*}} +1$, going as $a^{\alpha}$ for $a < a_{*}$ 
and approaching $1$ quite rapidly when $a > a_{*}$ and $\xi>1$. These profiles are illustrated 
in Fig. \ref{FIGU1} for $\xi = 2$ and for various values of $\alpha_{*}$ and $N_{*}$.
Different values of $\xi$ lead to  a transition which is more or less delayed; in any case, the explicit 
form of the profile only affects the power spectrum in a limited 
range of wavenumbers. The intermediate shaded area in Fig. \ref{FIGU1}
will therefore be immaterial for the accuracy of the present analysis.
The typical scale $a_{*}$ (roughly corresponding to the maximum of $n(a)$) 
may coincide with the end of the inflationary phase. 
The value of $a_{*}$ corresponds with a critical number of efolds $N_{*}$ 
which is of the order of $N_{t}$ (i.e. the total number of efolds)  if $a_{*}$ marks the end of the 
inflationary phase. This identification is however not mandatory and it will also be 
interesting to consider  the case $N_{*} < N_{t}$ or even $N_{*} \ll N_{t}$. 
 
It is finally useful to bear in mind that the scale factor during the different stages of the evolution 
are continuous and differentiable. An explicitly continuous 
parametrization of the scale factors is given by:
\begin{eqnarray}
a_{inf}(\tau) &=& \biggl(- \frac{\tau_{1}}{\tau}\biggr)^{\beta}, \qquad \tau \leq - \tau_{1},
\label{sfinf}\\
a_{r}(\tau) &=& \frac{\beta \tau + (\beta+1) \tau_{1}}{\tau_{1}}, \qquad -\tau_{1} < \tau \leq \tau_{2},
\label{sfrad}\\
a_{m}(\tau) &=& \frac{[ \beta \tau + \beta \tau_{2} + 2 (\beta+1) \tau_{1}]^2}{4 \tau_{1} [ \beta \tau_{2} + (\beta+ 1)\tau_{1}]}, \qquad \tau > \tau_{2},
\label{sfmat}
\end{eqnarray}
where $\tau_{1}$ coincides with the end of the inflationary phase and $\tau_{2}$ coincides with the time of matter-radiation equality; note that $\beta\to 1$ 
in the case of a pure de Sitter phase and $\beta = 1 - {\mathcal O}(\epsilon)$ in the quasi-de Sitter case.
The transition to the domination of the dark energy does not affect the slope and it has a mild 
effect on the amplitude of the spectrum \cite{dd1,dd2,absolute}; this effect will be included in the actual estimates of section 
\ref{sec4} by following the approach already pursued in \cite{absolute}.  A similar comment 
holds for other effects which are customarily associated with potential suppression of the amplitude of the power 
in the intermediate frequency branch such as the neutrino free streaming \cite{FS1,FS2} or the effective evolution of number of relativistic species. Equations (\ref{sfinf}), (\ref{sfrad}) and (\ref{sfmat}) are all continuous with their first derivatives at the transition points\footnote{This means, more specifically, that for  $\tau =-\tau_{1}$ we have that  $a_{i}(-\tau_{1}) = a_{r}(-\tau_{1})$ and $a_{i}'(-\tau_{1}) = a_{r}'(-\tau_{1})$. Similarly at the second transition point $a_{r}(\tau_{2}) = a_{m}(\tau_{2})$ and $a_{r}'(\tau_{2}) = a_{m}'(\tau_{2})$. }. If the scale factor is continuous at the transition points and if $n(a)$ 
is continuous and differentiable as a function of the scale factor, then the relation 
between $\eta$ and $\tau$ is also continuous and differentiable. All in all we can say that whenever $n(a)$ is a well behaved 
function interpolating between the two regimes illustrated in Fig. \ref{FIGU1} it will always be possible to consider 
the evolution as piecewise continuous without worrying about the evolution in the transition regime.

By then recalling Eq. (\ref{sixth}) and the explicit relation between $\eta$ and $\tau$, we have that for $a< a_{*}$  
the $\eta$-time and the conformal time are related as:
 \begin{equation}
 \eta = \int \frac{d\tau}{n_{*} a^{\alpha}(\tau)} = \frac{1}{n_{*}} \int d \tau \biggl(- \frac{\tau}{\tau_{*}}\biggr)^{\alpha\beta}.
 \label{MF2}
 \end{equation}
When $\tau_{*} = \tau_{1}$ (i.e. $N_{*} = N_{t}$) Eq. (\ref{MF2}) is explicitly given by
 \begin{equation}
 \biggl( - \frac{\eta}{\eta_{*}} \biggr) = \biggl( - \frac{\tau}{\tau_{*}}\biggr)^{\alpha\beta +1}, \qquad \frac{1}{\eta_{*}} = \frac{n_{*} (\alpha\beta +1)}{\tau_{*}},
 \label{MF3}
 \end{equation}
so that, in the conformal time parametrization,  $b(\tau)$ turns out to be:
\begin{equation}
b(\tau) = a(\tau) n^{\gamma -1/2}(\tau) = n_{*}^{\gamma -1/2} \biggl(- \frac{\tau}{\tau_{*}}\biggr)^{ - \beta[ 1 + \alpha (\gamma -1/2)]}.
\label{MF3a}
\end{equation} 
Using Eq. (\ref{MF3}) into Eq. (\ref{MF3a}) the expression of $b(\eta)$ is easily derived:
\begin{equation}
b(\eta) =  n_{*}^{\gamma -1/2} \biggl(- \frac{\eta}{\eta_{*}}\biggr)^{-\sigma}, \qquad \sigma = \frac{\beta[2 + \alpha ( 2 \gamma -1)]}{2 ( 1 + \alpha \beta)}. 
\label{MF4}
\end{equation}
In the regime of Eq. (\ref{MF4}) we can easily solve Eq. (\ref{Q10}) by defining the auxiliary mode function $F_{k}(\eta) b(\eta) = f_{k}(\eta)$.
The resulting equation can be solved in terms of Hankel functions and $F_{k}(\eta)$ is then given by\footnote{The solution (\ref{MF5}) can be expressed in the 
$\tau$-time by using Eq. (\ref{MF3}); depending on the convenience, 
both parametrizations can be consistently discussed. Note, finally,  that when $\beta = 1$, $\alpha =0$ and $\gamma =0$ the index $\mu$ equals $3/2$, as expected.}:
\begin{eqnarray}
F_{k}(\eta) &=& \frac{{\mathcal N}}{\sqrt{2 k} \,\,b(\eta)} \sqrt{- k \eta} \, \,H^{(1)}_{\mu}(- k \eta), \qquad {\mathcal N} = \sqrt{\frac{\pi}{2}} e^{i \, \pi(\sigma+ 1/2)/2},
\nonumber\\
\mu &=& \sigma + \frac{1}{2}=  \frac{1 + \beta[ ( 2 + \alpha) + \alpha( 2 \gamma -1)]}{2 ( 1 +\alpha \beta)},
\label{MF5}
\end{eqnarray}
where $H^{(1)}_{\mu}(k \eta)$ is the Hankel function\footnote{This solution reminds of the standard solution for the mode function; note, however, that in the 
present case the argument of the Hankel function 
is not $k \tau$ but rather $k\eta$. In terms of the $\tau$-parametrization the evolution is more complicated but equally solvable.} of the first kind \cite{abr1,abr2}.
Note that with the characteristic choice of the phase appearing in ${\mathcal N}$ the limit of Eq. (\ref{MF5}) for $\eta\to - \infty$ is exactly the one already mentioned 
in Eq. (\ref{Q10a}).

\subsection{Power spectra and tensor to scalar ratio}
Inserting Eqs. (\ref{MF4}) and (\ref{MF5})  into Eq. (\ref{MF1b}) 
the explicit form of the tensor power spectrum becomes:
\begin{equation}
{\mathcal P}_{T}(k,\eta) = \frac{k^2}{\pi \overline{M}_{P}^2} (- k \eta) \biggl(- \frac{\eta}{\eta_{*}}\biggr)^{2\sigma} n_{*}^{1 - 2 \gamma} \biggl|H_{\mu}^{(1)}(- k \eta)\biggr|^2.
\label{MF5a}
\end{equation}
In the simplest situation $\eta_{*} = \eta_{1}$ (i.e. $N_{*} = N_{t}$) and when the relevant modes are larger than the Hubble radius Eq. (\ref{MF5a}) can be modified by recalling Eq. (\ref{MF3}); in this limit we obtain \cite{abr1,abr2}:
\begin{equation}
\lim_{k\ll a H} {\mathcal P}_{T}(k,\eta)  = \frac{k^3\,\eta_{1}}{\pi^3\, \overline{M}_{P}^2}   \biggl(- \frac{\eta}{\eta_{1}}\biggr)^{2\sigma+1} 
n_{1}^{1 - 2 \gamma} \Gamma^2(\mu)\biggl(- \frac{k\eta}{2} \biggr)^{ - 2 \mu}.
\label{MF5b}
\end{equation}
Since the limits $k\ll a H$ and $k \eta \ll 1$ describe the same physical regime, after some algebra 
Eq. (\ref{MF5b}) becomes\footnote{The $\eta$-dependence in Eq. (\ref{MF5b}) cancels because $\mu= \sigma +1/2$, as it follows from Eq. (\ref{MF5}).}:
\begin{equation}
{\mathcal P}_{T}(k,\eta_{1}) =\biggl(\frac{H_{1}}{M_{P}}\biggr)^2 \frac{2^{ 2 \mu + 3}}{\pi^2} \Gamma^2(\mu)  n_{1}^{3 - 2 \gamma} \, \bigl|1 + \alpha\beta\bigr|^2 \, |k\eta_{1}|^{3 - 2 \mu}.
\label{MF5c}
\end{equation}
From Eq. (\ref{MF5c}) the tensor spectral index is defined as $n_{T} = 3 - 2 \mu$; using then Eq. (\ref{MF5}) the explicit expression 
of $n_{T}$ depends on $\alpha$, $\beta$ and $\gamma$ in the following manner:
\begin{eqnarray}
n_{T}(\alpha,\beta,\gamma) = \frac{2 ( 1 -\beta) + \alpha\beta( 3 - 2 \gamma)}{(1 + \alpha\beta)}.
\label{MF6}
\end{eqnarray}
The result of Eq. (\ref{MF5c}) can be further modified by including the slow-roll corrections\footnote{The slow-roll corrections are introduced in Eq. (\ref{MF6}) by appreciating that when $\epsilon \ll 1$ the value of $\beta$ becomes $\beta= 1/(1-\epsilon) = 1+ \epsilon$.} and by recalling, from  Eq. 
(\ref{MF3}), that $\eta_{1}$ and $\tau_{1}$ are related as $\eta_{1}= \tau_{1}/[n_{1} ( 1 + \alpha \beta)]$.  Equation (\ref{MF5c}) becomes:
\begin{equation}
{\mathcal P}_{T}(k,\tau_{1}) = \biggl(\frac{H_{1}}{M_{P}}\biggr)^2\,\, \frac{2^{6 - n_{T}}}{\pi^2} \,\Gamma^2\biggl(\frac{3 - n_{T}}{2}\biggr) \, n_{1}^{ 3 - n_{T} - 2\gamma}\, 
\biggl| 1 + \frac{\alpha}{1 - \epsilon}\biggr|^{2 -n_{T}} \,\,\biggl( \frac{k}{k_{\mathrm{max}}}\biggr)^{n_{T}}.
\label{MF8}
\end{equation}
Denoting with $\Omega_{{R}0}$ the present value of the critical fraction of radiative species (in the concordance paradigm 
photons and neutrinos) and with  ${\mathcal A}_{R}$ the amplitude of the scalar power spectrum at the pivot scale (see Eq. (\ref{ST1Ca})) the value of $k_{\mathrm{max}}$ in $\mathrm{Mpc}^{-1}$ units is:
\begin{equation}
\biggl(\frac{k_{\mathrm{max}}}{\mathrm{Mpc^{-1}}} \biggr) = 2.247 \times 10^{23}\, \biggl(\frac{H_{r}}{H_{1}}\biggr)^{\delta -1/2} \biggl(\frac{\epsilon}{0.01}\biggr)^{1/4} \biggl(\frac{{\mathcal A}_{{\mathcal R}}}{2.41 \times 10^{-9}} \biggr)^{1/4} \biggl(\frac{h_{0}^2 \Omega_{R0}}{4.15 \times 10^{-5}}\biggr)^{1/4}.
\label{MF9}
\end{equation}
In Eq. (\ref{MF9}) the $\delta$ accounts for the possibility of a delayed reheating terminating at an Hubble scale $H_{r}$
smaller than the Hubble rate during inflation. In what follows, as already mentioned in section \ref{sec1}, we shall 
rather stick to the conventional case where the reheating is sudden and $\delta = 1/2$ (or $H_{1} = H_{r}$ since the end of the 
inflationary phase coincides with the beginning of the radiation epoch
\footnote{In more general terms, however, 
 $H_{r}$ can be as low as $10^{-44} M_{P}$ (but not smaller) corresponding to a reheating scale occurring just prior to the formation of the light nuclei.}).

As suggested in Ref. \cite{two}, whenever $\alpha> 0$ the spectral index gets blue since it is always greater than the slow-roll 
correction. Using Eq. (\ref{MF6}) this statement can be made more precise by expanding 
 $n_{T}(\alpha,\beta, \gamma)$ in the limit $\epsilon <1$: 
\begin{eqnarray}
n_{T}(\alpha, \frac{1}{1-\epsilon}, \gamma)  &=& \frac{ \alpha ( 3 - 2 \gamma) - 2 \epsilon}{1 + \alpha - \epsilon}  
\label{MF10a}\\
&\to& 
\frac{\alpha ( 3 - 2 \gamma)}{1 + \alpha} + \frac{[ -2 + \alpha ( 1 - 2\gamma)]\epsilon}{(1 + \alpha)^2} + {\mathcal O}(\epsilon^2).
\label{MF10b}
\end{eqnarray}
When $\gamma=0$ we are in the situation of Ref. \cite{two} and  Eqs. (\ref{MF10a})--(\ref{MF10b}) imply that the tensor spectral index is: 
\begin{equation}
n_{T}  = \frac{3 \alpha}{1 + \alpha} + \frac{[ -2 + \alpha ]\epsilon}{(1 + \alpha)^2} + {\mathcal O}(\epsilon^2).
\label{MF11}
\end{equation}
When $\gamma=1$ we are in the situation of Ref. \cite{three} and Eqs. (\ref{MF10a})--(\ref{MF10b}) implies instead:
\begin{equation}
n_{T}  = \frac{\alpha}{1 + \alpha} - \frac{(2 + \alpha )\epsilon}{(1 + \alpha)^2} + {\mathcal O}(\epsilon^2).
\label{MF12}
\end{equation}
Finally when $\alpha = \gamma =0$ the standard result is recovered, i.e. $n_{T} = - 2 \epsilon + {\mathcal O}(\epsilon^2)$.
Concerning the results of Eqs. (\ref{MF10a})--(\ref{MF10b}) and of Eqs. (\ref{MF11})--(\ref{MF12}) few comments are in order.
From  Eq. (\ref{MF10a}) it is immediate to appreciate that the spectrum is blue and $n_{T}>0$ provided:
\begin{eqnarray}
\epsilon < 1 + \alpha, \qquad \epsilon < \frac{\alpha ( 3 - 2\gamma)}{2};
\label{MF12a}
\end{eqnarray}
these two conditions are always verified in practice by definition 
of slow-roll parameter (implying $\epsilon \ll 1$). 
All in all we can then say that within the physical range of the parameters the presence of $\gamma$ just shifts 
the value of the tensor spectral index (see Eqs. (\ref{MF11}) and (\ref{MF12})) but does not 
change qualitatively the whole picture\footnote{From  Eq. (\ref{MF10b}) we have that $n_{T} > 0$ iff $\gamma < 3/2$: it is 
interesting to appreciate that large values of $\gamma$ lead to a red spectrum so that, even in the
heuristic approach mentioned after Eq. (\ref{fifth}) (which is not the one followed here) we should always demand $0\leq \gamma < 3/2$.}.

If the transition to normalcy
is sufficiently sudden, the details of the transition will be immaterial for the large-scale 
spectrum. We then have that the tensor to scalar ratio is given by\footnote{
Denoting with $\eta_{i}$ the initial time of the evolution (for instance at the onset of inflation) we shall have that
$n_{i} = n_{1} (a_{i}/a_{1})^ {\alpha} = n_{1} e^{ - \alpha N_{*}}$ with  $n_{i} \geq 1$.
When $n_{i} =1$, $n_{1} \to \exp{(\alpha N_{*})}$ (see also Fig. \ref{FIGU1} and discussion therein).}: 
\begin{equation}
r_{T}(k_{p}) = \epsilon \frac{2^{6 - n_{T}}}{\pi} \Gamma^2\biggl(\frac{3 - n_{T}}{2} \biggr) n_{i}^{3 - n_{T} - 2 \gamma} e^{N_{*} \alpha ( 3 - n_{T} - 2 \gamma)}\
\biggl| 1 + \frac{\alpha}{ 1 - \epsilon}\biggr|^{ 2 - n_{T}} \biggl(\frac{k_{p}}{k_{\mathrm{max}}}\biggr)^{n_{T}}.
\label{MF15}
\end{equation}
Equations  (\ref{MF10a}) and (\ref{MF15}) demonstrate that the consistency relations do not hold in the case 
where the gravitons have a refractive index. A similar phenomenon 
has been suggested in \cite{violation} as a consequence of the opposite physical situation where 
the initial thermal bath of primordial phonons (i.e. large-scale quanta of curvature perturbations) 
is characterized by a sound speed different from the speed of light. 

The large-scale power spectra are not modified in the large-scale limit as long as the evolution 
of $b(\eta)$ is continuous. For this purpose, without indulging in the explicit analytic expression 
in the transition region, we can consider the situation where before $-\eta_{*}$ the large-scale power spectrum is given 
by the solution (\ref{MF5}). The resulting power spectrum at a later time is therefore directly computable 
since, in the large-scale limit, Eq. (\ref{Q10}) becomes\footnote{Note that in the large-scale limit the term $k^2 F_{k}(\eta)$ is negligible in Eq. (\ref{Q10}).} 
 $\partial_{\eta}[ b^2(\eta) \partial_{\eta} F_{k}(\eta)]= 0$.
This equation can then be explicitly integrated from $-\eta_{2}$ 
onwards, namely:
\begin{equation}
F_{k}(\eta) = F_{k}(-\eta_{2}) + G_{k}(-\eta_{2}) \int_{- \eta_{2}}^{\eta} \frac{d\eta^{\prime}}{b^2(\eta^{\prime})}, \qquad 
 G_{k}(-\eta_{2}) = b^2(- \eta_{2}) \dot{F}_{k}(-\eta_{2}),
\label{MF16}
\end{equation}
where the time $-\eta_{2}$ is a generic reference time before $-\eta_{*}$ and very close to it. 
With this observation we have that the power spectrum after $ - \eta_{*}$ can be written 
as 
\begin{equation}
{\mathcal P}_{T}(k, \eta) = \frac{4 \ell_{P}^2 k^3}{\pi^2} \bigl| F_{k}(-\eta_{*})\bigr|^2 \biggl[ 1 + \biggl(\frac{1}{2} - \mu\biggr) b_{*}^2\int_{-\eta_{*}}^{\eta} \frac{ d (\eta^{\prime}/\eta_{*})}{b^2(\eta^{\prime})} \biggr]^2.
\label{MF17}
\end{equation}
Equation (\ref{MF17}) shows that the only thing that counts is the continuity of $b(\eta)$ across the transition.
Well after $-\eta_{*}$ we also have that $b(\eta)\to a(\eta) = a(\tau)$ and the background still evolves 
like quasi-de Sitter with slow-roll parameter given by $\epsilon$. It is then clear that the second term inside 
the square bracket is subleading and this shows, as expected, that the large-scale power spectrum 
at large-scale is not affected by this kind of continuous transitions. 

\subsection{The spectral energy density}
The independent variation of the tensor spectral index $n_{T}$ and of the tensor to scalar 
ratio $r_{T}(k_{p})$ can be investigated from the absolute normalization of the 
spectral energy density \cite{absolute}. 
When the relevant modes are today inside the Hubble radius\footnote{The second term inside the square bracket of Eq. (\ref{PH1}) is actually subleading when the relevant scales are larger than the Hubble radius, i.e.
$k\tau\gg 1$.} the relation 
between the spectral energy density and the power spectrum at the corresponding 
time is given by:
\begin{equation}
\Omega_{\mathrm{gw}}(k,\tau) = \frac{k^2}{12 H^2 a^2} {\mathcal P}_{\mathrm{T}}(k,\tau)\biggl[ 1 + {\mathcal O}\biggl(\frac{{\mathcal H}^2}{k^2}\biggr)\biggr], \qquad {\mathcal H} = \frac{a^{\prime}}{a}.
\label{PH1}
\end{equation}
There are two possibilities for an accurate assessment of the absolute normalization appearing in Eq. (\ref{PH1}). In the first approach we 
can compute the transfer function for the amplitude as \cite{second,absolute,absolute0,absolute2,absolute3}
\begin{eqnarray}
T_{h}(k/k_{\mathrm{eq}}) &=& \sqrt{1 + c_{1} \biggl(\frac{k}{k_{\mathrm{eq}}}\biggr) + b_{1} \biggl(\frac{k}{k_{\mathrm{eq}}}\biggr)^2},\qquad 
c_{1}= 1.260, \qquad b_{1}= 2.683,
\label{PH2}\\
k_{\mathrm{eq}} &=& 0.0732\,\, h_{0}^2 \Omega_{\mathrm{M}0} \, \,\mathrm{Mpc}^{-1}
\label{PH2a}
\end{eqnarray}
where $\Omega_{\mathrm{M}0}$ denotes the present critical fraction of dusty matter in the concordance paradigm.
 By using the transfer function for the tensor amplitude, 
the spectral energy density for frequencies $\nu \gg \nu_{\mathrm{eq}}$ can be simply given by:
\begin{eqnarray}
h_{0}^2 \Omega_{\mathrm{gw}}(\nu,\tau_{0}) &=& {\mathcal N}_{h} \,\, r_{\mathrm{T}} \,\, \biggl(\frac{\nu}{\nu_{\mathrm{p}}}\biggr)^{n_{\mathrm{T}}} e^{- 2\beta \frac{\nu}{\nu_{\mathrm{max}}}},
\label{PH3}\\
{\mathcal N}_{h} &=& 7.992 \times 10^{-15} \biggl(\frac{h_{0}^2 \Omega_{\mathrm{M}0}}{0.1326}\biggr)^{-2} 
\biggl(\frac{h_{0}^2 \Omega_{\mathrm{R}0}}{4.15\times 10^{-5}}\biggr) \biggl(\frac{d_{\mathrm{A}}}{1.4115 \times 10^{4}\, \mathrm{Mpc}}\biggr)^{-4},
\label{PH4}
\end{eqnarray}
where $d_{\mathrm{A}}$ is the (comoving) angular diameter distance to decoupling and $\Omega_{\mathrm{R}0}$ is the fraction of critical energy density attributed to massless particles (photons and neutrinos in the concordance paradigm). 
The second strategy is to use the direct integration of the mode functions and discuss the transfer function directly in terms of the spectral energy density \cite{absolute,absolute2}:
\begin{eqnarray}
h_{0}^2 \Omega_{\mathrm{gw}}(\nu,\tau_{0}) &=& {\mathcal N}_{\rho}  T^2_{\rho}(\nu/\nu_{\mathrm{eq}}) r_{\mathrm{T}}(k_{p}) \biggl(\frac{\nu}{\nu_{\mathrm{p}}}\biggr)^{n_{\mathrm{T}}} e^{- 2 \beta \frac{\nu}{\nu_{\mathrm{max}}}}
\label{PH5}\\
{\mathcal N}_{\rho} &=& 4.165 \times 10^{-15} \biggl(\frac{h_{0}^2 \Omega_{\mathrm{R}0}}{4.15\times 10^{-5}}\biggr).
\label{PH6}
\end{eqnarray}
where $\nu_{\mathrm{eq}}$ denotes the equality frequency:
\begin{equation}
\nu_{\mathrm{eq}} =  \frac{k_{\mathrm{eq}}}{2 \pi} = 1.597\times 10^{-17} \biggl(\frac{h_{0}^2 \Omega_{\mathrm{M}0}}{0.1411}\biggr) \biggl(\frac{h_{0}^2 \Omega_{\mathrm{R}0}}{4.15 \times 10^{-5}}\biggr)^{-1/2}\,\, \mathrm{Hz},
\label{eight}
\end{equation}
where  
$T_{\rho}(\nu/\nu_{\mathrm{eq}})$ is the transfer function of the spectral energy density:
\begin{equation}
T_{\rho}(\nu/\nu_{\mathrm{eq}}) = \sqrt{1 + c_{2}\biggl(\frac{\nu_{\mathrm{eq}}}{\nu}\biggr) + b_{2}\biggl(\frac{\nu_{\mathrm{eq}}}{\nu}\biggr)^2},\qquad c_{2}= 0.5238,\qquad
b_{2}=0.3537.
\label{EQ19}
\end{equation}
Equation (\ref{EQ19}) is obtained by integrating the mode functions across the radiation-matter transition and by computing 
$\Omega_{\mathrm{gw}}(\nu,\tau)$ for different frequencies.  By comparing Eqs. (\ref{PH5})--(\ref{PH6}) to Eqs. (\ref{PH3})--(\ref{PH4}), the amplitude for $\nu\gg \nu_{\mathrm{eq}}$ differs, roughly,  
by a factor $2$. This coincidence is not surprising since  Eqs. (\ref{PH3})--(\ref{PH4}) have been obtained by averaging over the 
oscillations (i.e. by replacing cosine squared with $1/2$).  These manipulations are less accurate than the procedure used to derive the transfer function for the spectral energy density, as stressed in the past \cite{absolute}.

Recalling Fig. \ref{FIGU1} and the related discussion, when the critical number of efolds equals the total number of efolds (i.e. $N_{*} = N_{t}$)
 the form of the spectral energy density is the same as 
in the standard case with the difference that the tensor to scalar ratio and the spectral index dapend on the 
parameters of the model and are given, respectively, by Eqs. (\ref{MF6}) and (\ref{MF15}). If $N_{*} < N_{t}$  the spectral energy density is characterized by two branches. More specifically the full expression of the spectral energy density can be written as:
\begin{eqnarray} 
h_{0}^2 \Omega_{\mathrm{gw}}(\nu,\tau_{0}) &=& {\mathcal N}_{\rho}  T^2_{\rho}(\nu/\nu_{\mathrm{eq}}) r_{\mathrm{T}}(\nu_{p}) \biggl(\frac{\nu}{\nu_{\mathrm{p}}}\biggr)^{n_{\mathrm{T}}}, \qquad \nu < \nu_{*}
\label{NN1}\\
&=&  {\mathcal N}_{\rho} r_{\mathrm{T}}(\nu_{p}) \biggl(\frac{\nu_{*}}{\nu_{\mathrm{p}}}\biggr)^{n_{\mathrm{T}}}\,\biggl(\frac{\nu}{\nu_{*}}\biggr)^{- 2 \epsilon}\, e^{- 2 \beta \frac{\nu}{\nu_{\mathrm{max}}}}, \qquad \nu > \nu_{*},
\label{NN2}
\end{eqnarray} 
where $\nu_{*}$ and $\nu_{\mathrm{max}}$ are given, respectively, by:
\begin{eqnarray}
\nu_{\mathrm{max}} &=& 0.3 \,\biggl(\frac{r_{T}}{0.1}\biggr)^{1/4} 
\biggl(\frac{{\mathcal A}_{\mathcal R}}{2.41 \times 10^{-9}}\biggr)^{1/4}
\biggl(\frac{h_{0}^2 \Omega_{\mathrm{R}0}}{4.15 \times 10^{-5}}\biggr)^{1/4} \,\mathrm{GHz},
\label{CC4}\\
\nu_{*} &=& \biggl(1 + \frac{\alpha}{1 - \epsilon}\biggr) e^{N_{*}(\alpha + 1) - N_{t}}\nu_{\mathrm{max}}.
\label{nustar}
\end{eqnarray}
In general terms, as we shall see, $\nu_{*}$ will always be larger than ${\mathcal O}(10^{-11})$ Hz but,
depending on the values of $\alpha$ and $N_{*}$ it can be either larger or smaller than the Hz. 

\renewcommand{\theequation}{4.\arabic{equation}}
\setcounter{equation}{0}
\section{Phenomenological implications}
\label{sec4}
The rate of variation of the refractive index  $\alpha$ and the critical number of efolds $N_{*}$ are 
the  two pivotal parameters for the analysis reported in the present section.
If $N_{*}$ is of the order of $N_{t}$ the transition to normalcy occurs at the end of inflation;
conversely when $N_{*} <N_{t}$ (as in the case of Fig. \ref{FIGU1})
the transition to normalcy takes place well before the onset of the radiation-dominated epoch
(i.e. when the background is still inflating deep inside the quasi-de Sitter stage of expansion).
These two different situations will now be separately examined by assuming, for illustrative 
purposes, a fiducial number of inflationary efolds $N_{t} = {\mathcal O}(70)$.
Larger values of $N_{t}$ do not affect the main conclusions inferred below.

\subsection{Basic constraints}
The scale-dependent constraint on the tensor to scalar ratio is 
 conventionally applied at the pivot scale $k_{p}= 0.002\,\, \mathrm{Mpc}^{-1}$ corresponding 
to the comoving frequency $\nu_{p} = k_{p}/(2 \pi)$. While the Planck data alone 
would suggest an upper limit $r_{T}(k_{p}) < 0.11$, the joint analysis of Planck and BICEP2/Keck array data \cite{BICPL}
already demands $r_{T}(k_{p})<0.07$. In a conservative perspective we shall therefore require the limit: 
\begin{equation}
r_{T}(\nu_{p}) \leq 0.06, \qquad \nu_{p} = \frac{k_{p}}{2 \pi} = 3.092\times 10^{-18} \,\, \mathrm{Hz}= 3.092\,\,\mathrm{aHz}
\label{CC1}
\end{equation}
on the tensor to scalar ratio computed in Eq. (\ref{MF15}).
\begin{figure}[!ht]
\centering
\includegraphics[height=6.5cm]{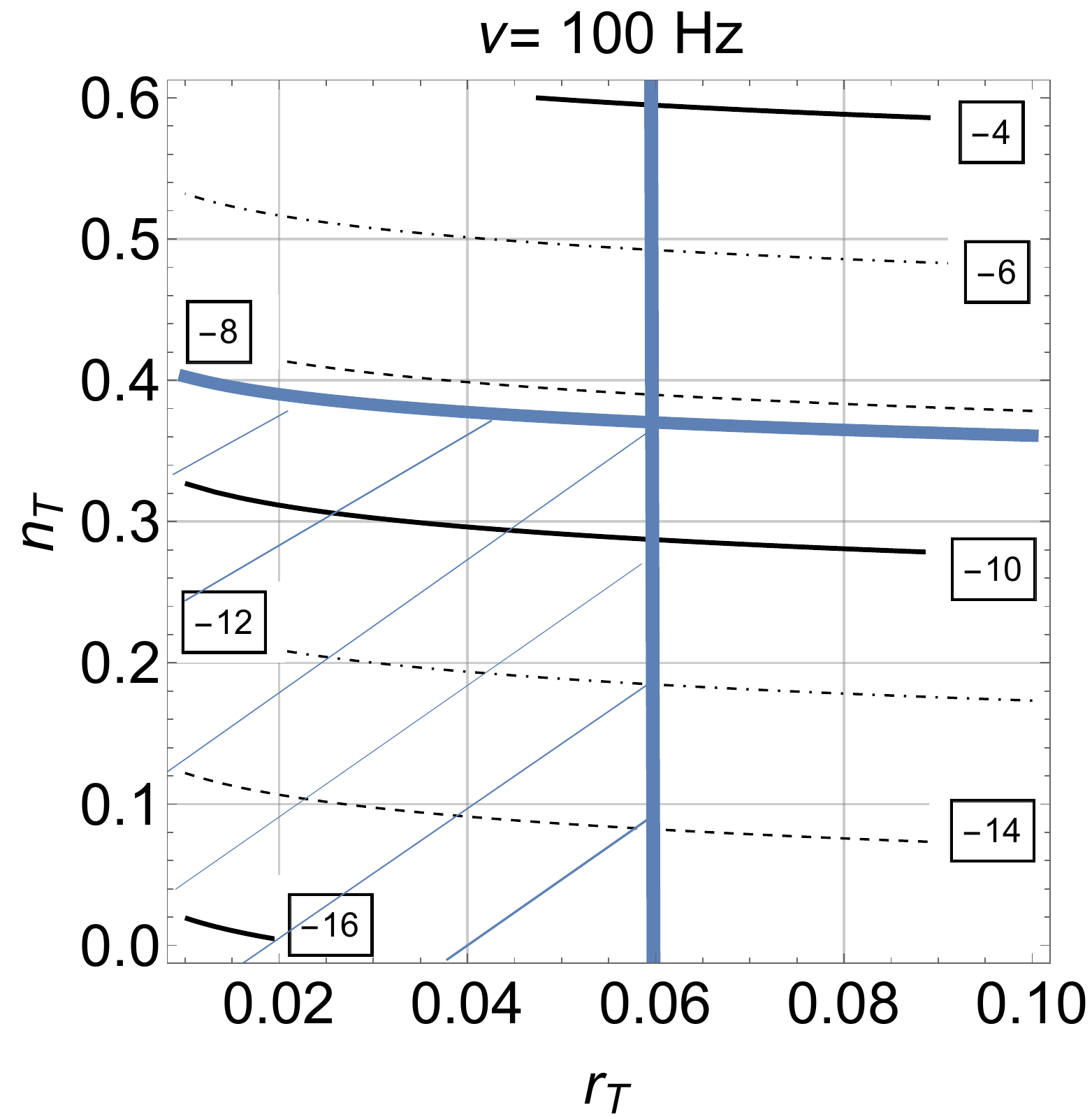}
\includegraphics[height=6.5cm]{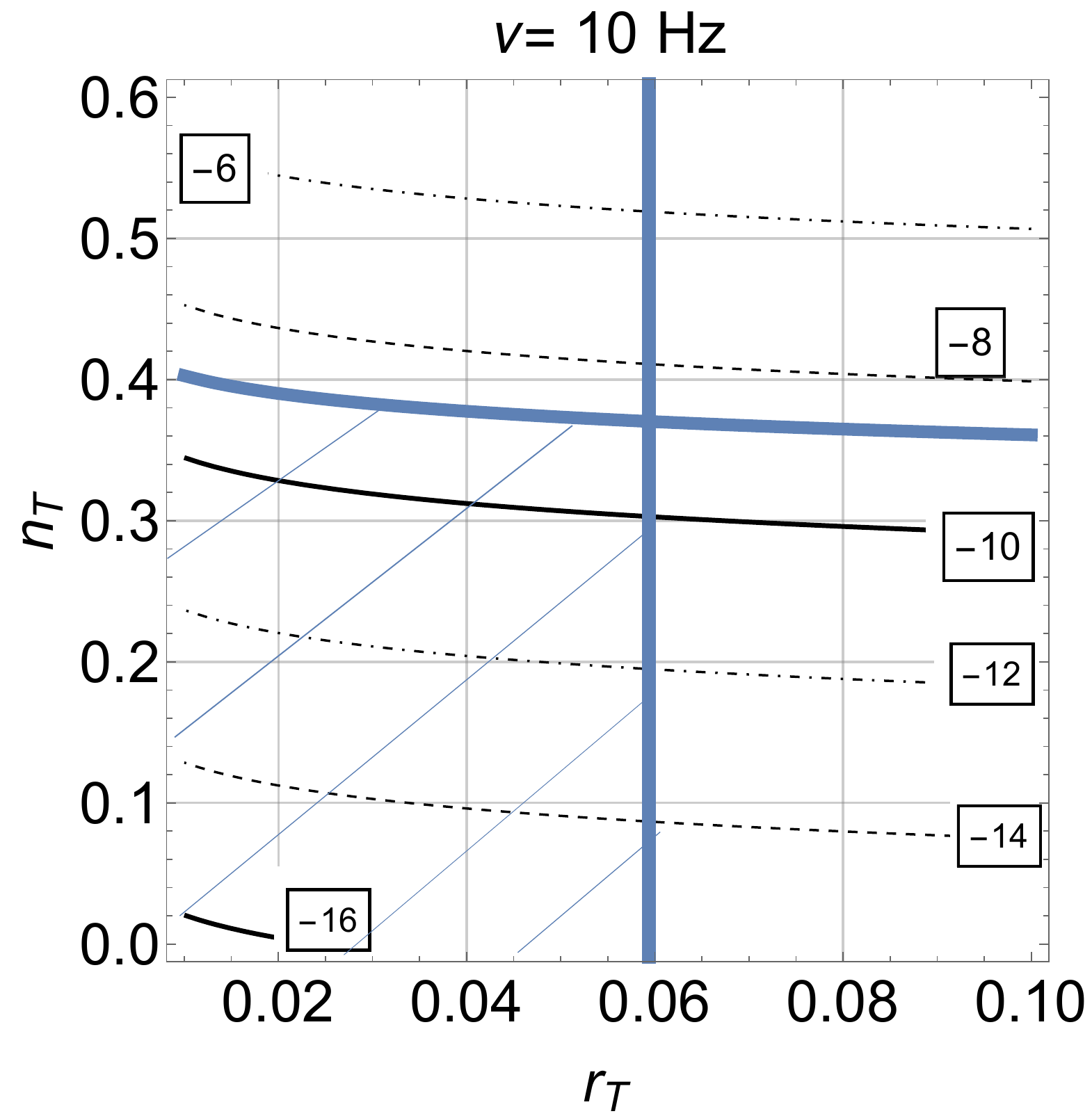}
\caption[a]{In both plots the various curves and the corresponding labels  illustrate different values of the common logarithm of $h_{0}^2 \Omega_{\mathrm{gw}}$. The two plots refer to two different frequencies within the audio band.}
\label{FIGU2}      
\end{figure}
The bounds coming from big-bang nucleosynthesis \cite{bbn1,bbn2,bbn3} imply instead a constraint
on the integral of the spectral energy density of the gravitons:
\begin{equation}
h_{0}^2  \int_{\nu_{\mathrm{bbn}}}^{\nu_{\mathrm{max}}}
  \Omega_{\mathrm{gw}}(\nu,\tau_{0}) d\ln{\nu} = 5.61 \times 10^{-6} \Delta N_{\nu} 
  \biggl(\frac{h_{0}^2 \Omega_{\gamma0}}{2.47 \times 10^{-5}}\biggr).
\label{CC2}
\end{equation}
The lower limit of integration in Eq. (\ref{CC2}) is given by  the frequency corresponding to the Hubble rate at the nucleosynthesis epoch: 
\begin{equation}
\nu_{\mathrm{bbn}}= 2.252\times 10^{-11} \biggl(\frac{g_{\rho}}{10.75}\biggr)^{1/4} \biggl(\frac{T_{\mathrm{bbn}}}{\,\,\mathrm{MeV}}\biggr) 
\biggl(\frac{h_{0}^2 \Omega_{\mathrm{R}0}}{4.15 \times 10^{-5}}\biggr)^{1/4}\,\,\mathrm{Hz} \simeq 0.01\, \mathrm{nHz},
\label{CC3}
\end{equation}
where  $g_{\rho}$ denotes the effective number of relativistic degrees of freedom entering the total energy density of the plasma and $T_{\mathrm{bbn}}$ is the temperature of big-bang nucleosynthesis. The limit  of Eq. (\ref{CC2}) sets an indirect constraint  on the extra-relativistic species (and, among others, on the relic gravitons). Since Eq. (\ref{CC2}) is relevant in the context of neutrino physics (when applied to massless fermionic species), the limit is often expressed for practical reasons  in terms of $\Delta N_{\nu}$ representing the contribution of supplementary 
neutrino species. The actual bounds on $\Delta N_{\nu}$ range from $\Delta N_{\nu} \leq 0.2$ 
to $\Delta N_{\nu} \leq 1$;  the integrated spectral density in Eq. (\ref{CC2}) is thus between $10^{-6}$ and $10^{-5}$. 
Finally the pulsar timing measurements determine a local bound on the spectral energy density \cite{PUL1,PUL2}
\begin{equation}
\Omega_{\mathrm{gw}}(\nu_{pul},\tau_{0}) < 1.9 \times10^{-8}, \qquad \nu_{pul} \simeq \,10^{-8}\,\mathrm{Hz},
\label{CC1a}
\end{equation}
which is an upper limit at a typical frequency, roughly corresponding to the inverse of the observation 
time along which the timing has been monitored. Various applications and refinements of the pulsar timing cosntraint 
have been discussed through the years and also recently (see e.g. Refs. \cite{PUL3,PUL4,PUL5,PUL6}).

\subsection{Target sensitivities and speculations}
\begin{figure}[!ht]
\centering
\includegraphics[height=6.5cm]{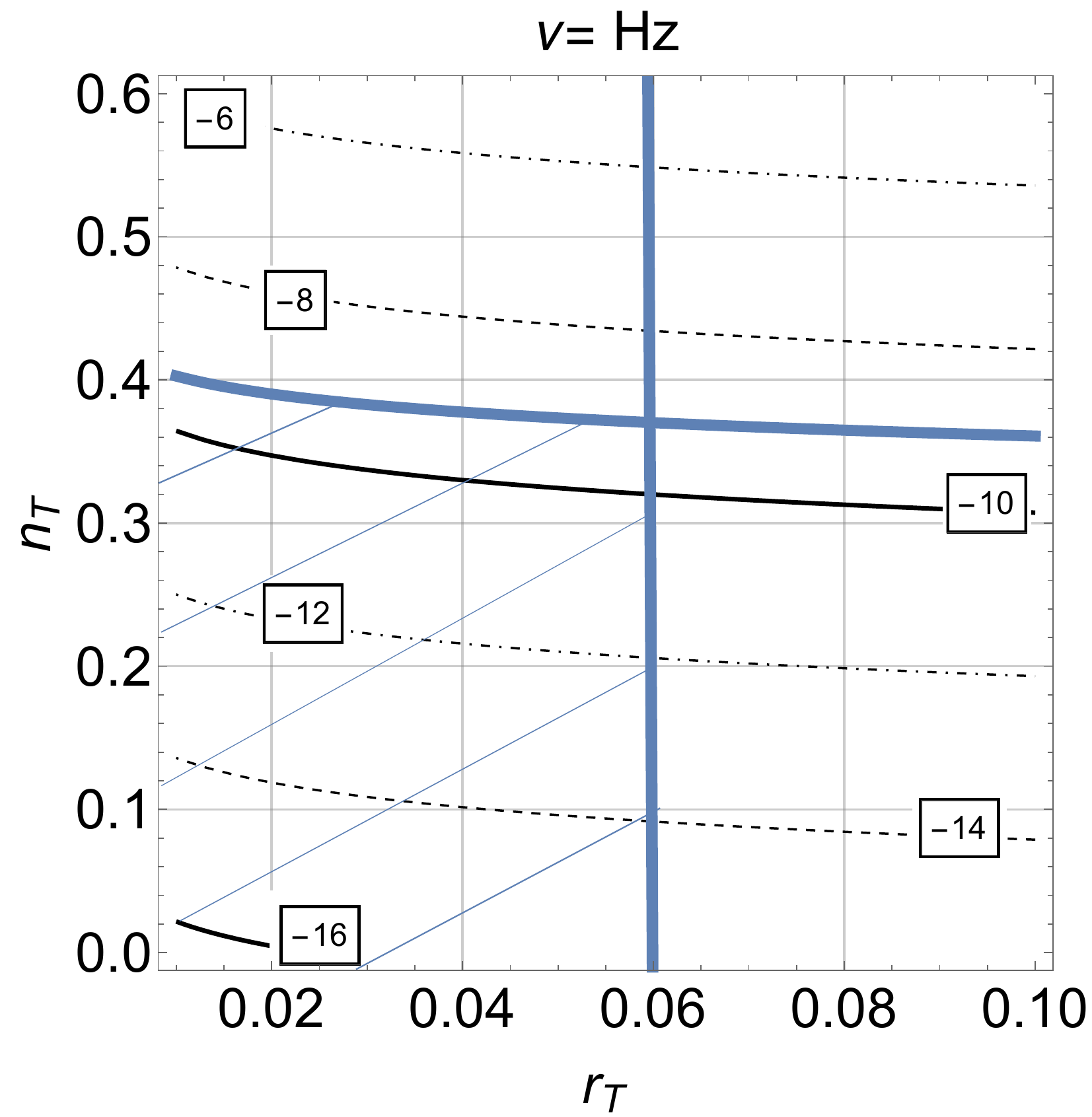}
\includegraphics[height=6.5cm]{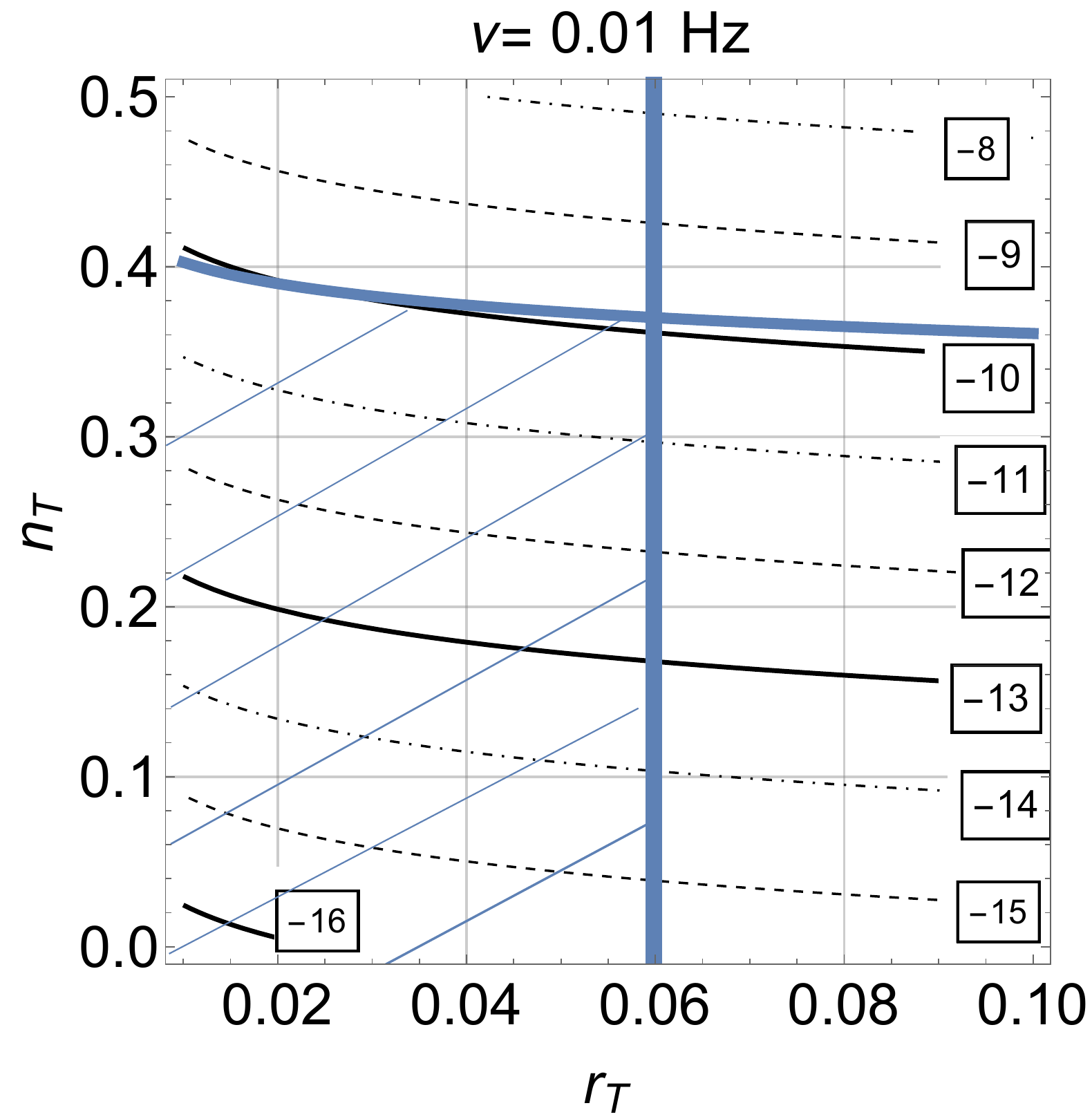}
\caption[a]{As in Fig. \ref{FIGU2} the curves illustrate the different values of the common logarithm of $h_{0}^2 \Omega_{\mathrm{gw}}$ reported 
in the corresponding labels. The frequencies of the two plots fall within the mHz band.}
\label{FIGU3}      
\end{figure}

Generally speaking the phenomenological constraints of Eqs. (\ref{CC1}), (\ref{CC2} and (\ref{CC1a})
are satisfied in a certain corner of the parameter space of a given model. The game will then be to see 
to what extent these regions overlap with the areas where the signal is sufficiently 
large to be detected in the future. For instance in Figs. \ref{FIGU1} and \ref{FIGU2} 
the constraints of Eqs. (\ref{CC1}), (\ref{CC2}) and (\ref{CC1a})
are satisfied in the shaded area which has been obtained by assuming 
(incorrectly, as we shall see in a moment) that the spectral index and the tensor to scalar ratio change independently. 
The labels appearing on the curves of the plots 
of Figs. \ref{FIGU2} and \ref{FIGU3} denote the common logarithm of $h_{0}^2 \Omega_{\mathrm{gw}}$ 
at the four frequencies indicated above each of the plots. 

The two frequencies discussed in Fig. \ref{FIGU2} 
correspond, respectively, to $10$ and $100$ Hz: they illustrate the spectral energy density in the 
so-called {\em audio band}  ranging between few Hz and $10$ kHz. Terrestrial wide-band 
interferometers, depending on their respective physical dimensions, operate within this band.
The two frequencies illustrated in Fig. \ref{FIGU3} fall instead within the {\em mHz} band 
ranging between a fraction of the mHz and few Hz. Space-borne detectors operate 
within this band. We can finally define a {\em MHz band}, extending between $100$ kHz and few GHz;
this band is irrelevant for the signal coming from conventional inflationary models but could play 
a relevant role in the present context since, in this case, most of the signal is concentrated exactly 
in this region.

While the current upper limits on stochastic backgrounds of relic gravitons 
are still far from the final targets \cite{LV3,LV4}, the advanced Ligo/Virgo projects are described in \cite{LV1,LV2}. 
We shall then suppose, according to Refs. \cite{LV1,LV2}  that the terrestrial interferometers (in their 
advanced version) will be one day able to probe chirp amplitudes ${\mathcal O}(10^{-25})$ corresponding 
to spectral amplitudes $h_{0}^2 \Omega_{\mathrm{gw}} = {\mathcal O}(10^{-11})$.  
In the foreseeable future there should be two further interferometers (with different designs) operational 
in the audio band namely the Japanese Kagra\footnote{Kagra is somehow the prosecution of the Tama-300 experiment \cite{TAMA}). Among the 
wide-band detectors we should also mention the GEO-600 experiment \cite{GEO1} which is 
now included in the Ligo/Virgo consortium \cite{GEO2}. } (Kamioka Gravitational Wave Detector)
\cite{kagra1,kagra2} and the Einstein telescope \cite{ET1} both with improved 
sensitivities in comparison with the advanced Ligo/Virgo interferometers.
 
The target sensitivity to detect the stochastic background of inflationary origin 
should be $h_{c}={\mathcal O}(10^{-29})$ (or smaller) in the chirp amplitude and 
$h_{0}^2 \Omega_{\mathrm{gw}} = {\mathcal O}(10^{-16})$ (or smaller) 
in the spectral energy density. These figures directly come from the amplitude of the quasi-flat plateau 
produced in the context of single-field inflationary scenarios; in this case the plateau
encompasses the mHz and the audio bands with basically the same amplitude. 
Even though these sensitivities are beyond reach for the current interferometers, 
a number of ambitious projects will be hopefully operational in the future.
The  space-borne interferometers, such as (e)Lisa (Laser Interferometer Space Antenna) \cite{LISA}, Bbo (Big Bang Observer) 
\cite{BBO}, and Decigo (Deci-hertz Interferometer Gravitational Wave Observatory) \cite{DECIGO1,DECIGO2}, 
might operate between few mHz and the Hz hopefully within the following score year. While the sensitivities of these 
instruments are still at the level of targets, we can say that they should probably range 
between $h_{0}^2 \Omega_{\mathrm{gw}} = {\mathcal O}(10^{-12})$ and $h_{0}^2 \Omega_{\mathrm{gw}} = {\mathcal O}(10^{-15})$.

In spite of the minute signals related to conventional inflationary scenarios, 
if the spectrum increases in frequency (i.e. $n_{T} >0$) Figs. \ref{FIGU2} and \ref{FIGU3}
seem to suggest  potentially large signals 
both for terrestrial interferometers (in their advanced version) and for 
space-borne detectors. However, this way of reasoning is misleading, at least partially.
Indeed, while it is true that in the present situation the consistency relations are violated, 
it is equally true that $r_{T}$ and $n_{T}$ depend on $\epsilon$, $N_{*}$, $\alpha$ and $\gamma$ 
in a specific way (see e.g. Eqs. (\ref{MF10a}) and (\ref{MF15})). 
\begin{figure}[!ht]
\centering
\includegraphics[height=6.5cm]{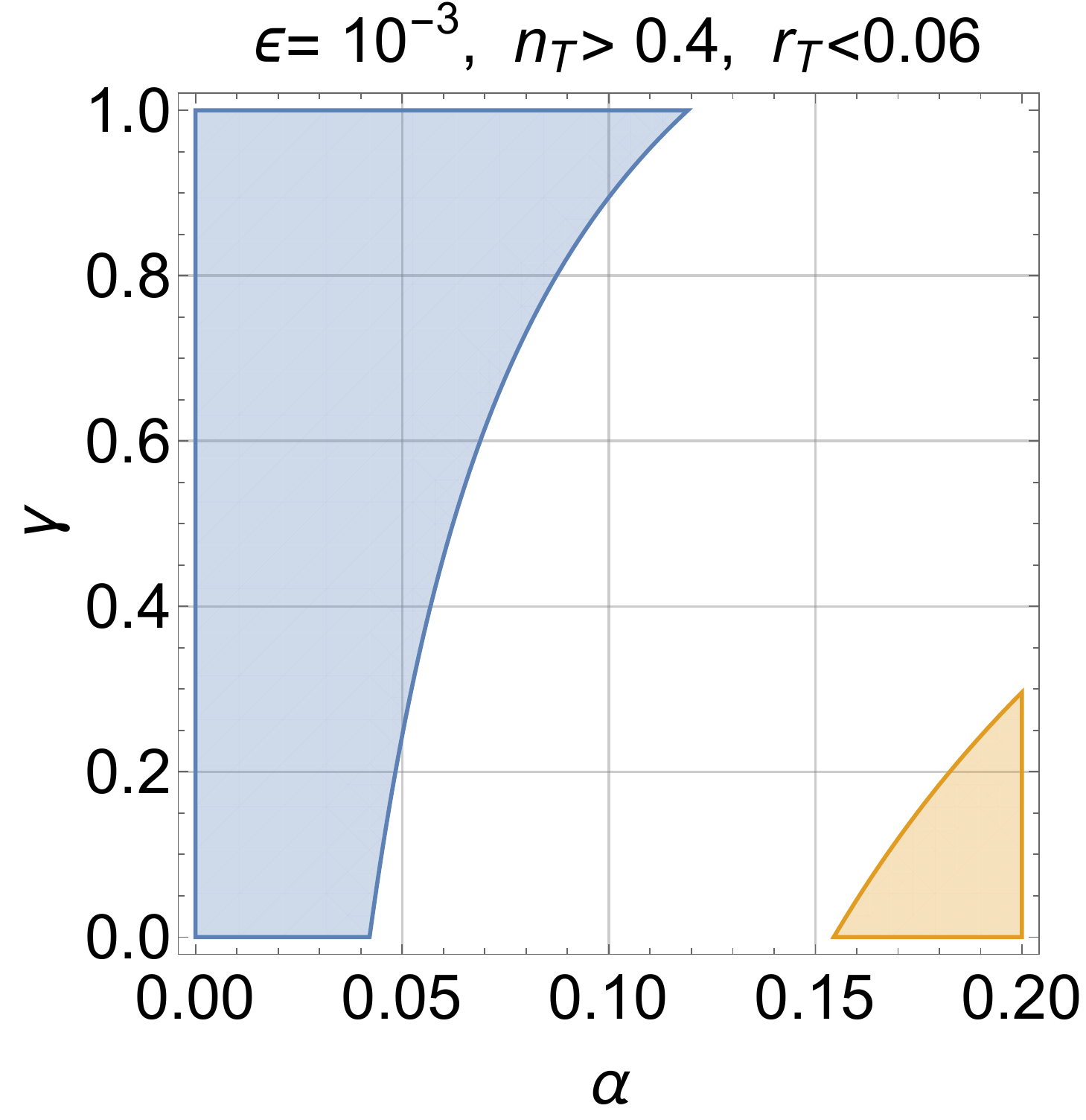}
\includegraphics[height=6.5cm]{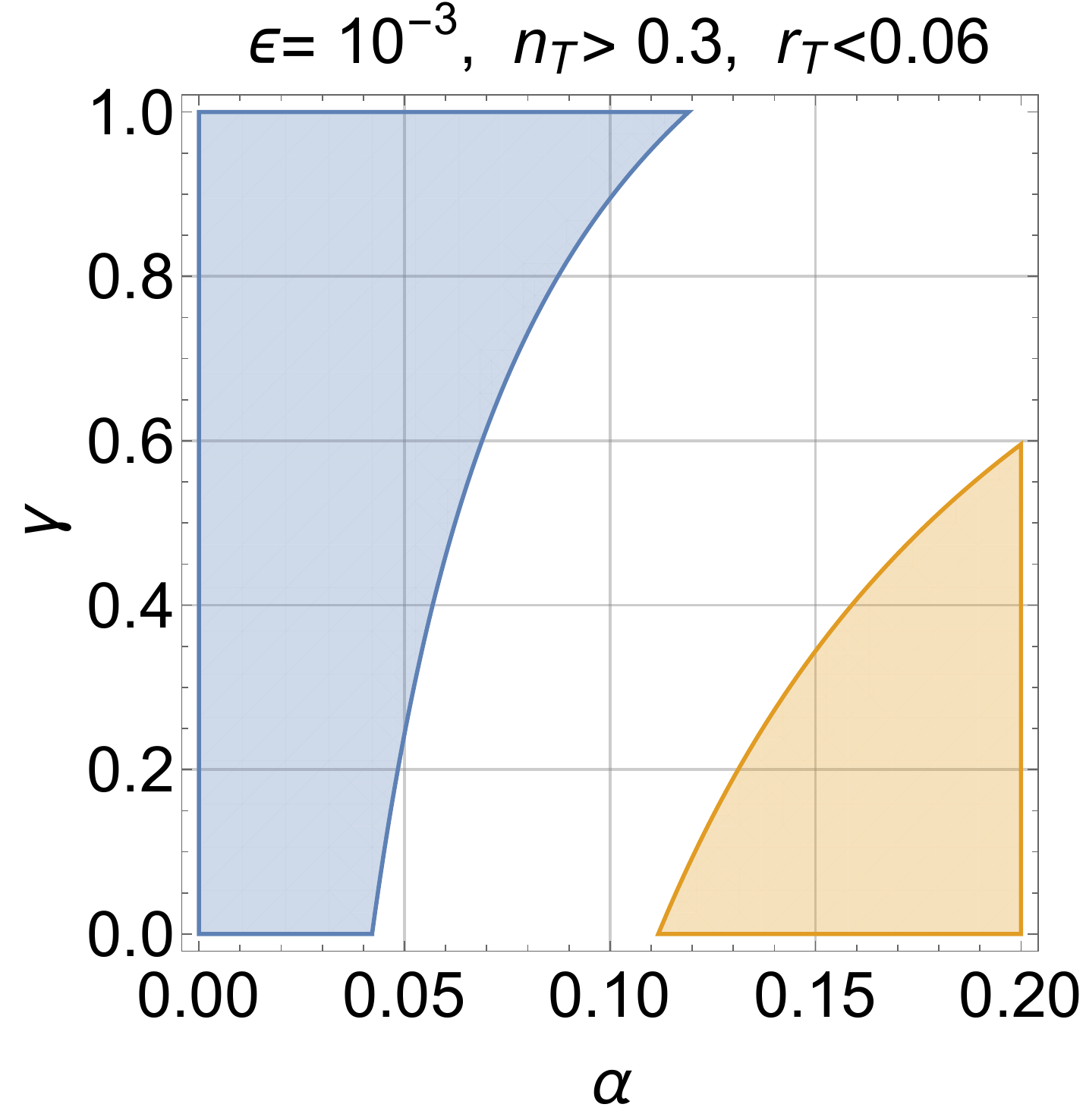}
\caption[a]{The constraints on $r_{T}$ and $n_{T}$ are translated in the $(\alpha,\,\gamma)$ plane for the values of $n_{T}$ 
compatible with an observable signal in the audio band (i.e. $\mathrm{Hz} \leq \nu \leq 10^{4} \mathrm{Hz}$) and in the generic case 
of an instrument able to probe spectral energy densities  ${\mathcal O}(10^{-10})$.}
\label{FIGU4}      
\end{figure}

\subsection{Charting the parameter space in the $ N_{*} = {\mathcal O}(N_{t})$ case} 
\begin{figure}[!ht]
\centering
\includegraphics[height=6.5cm]{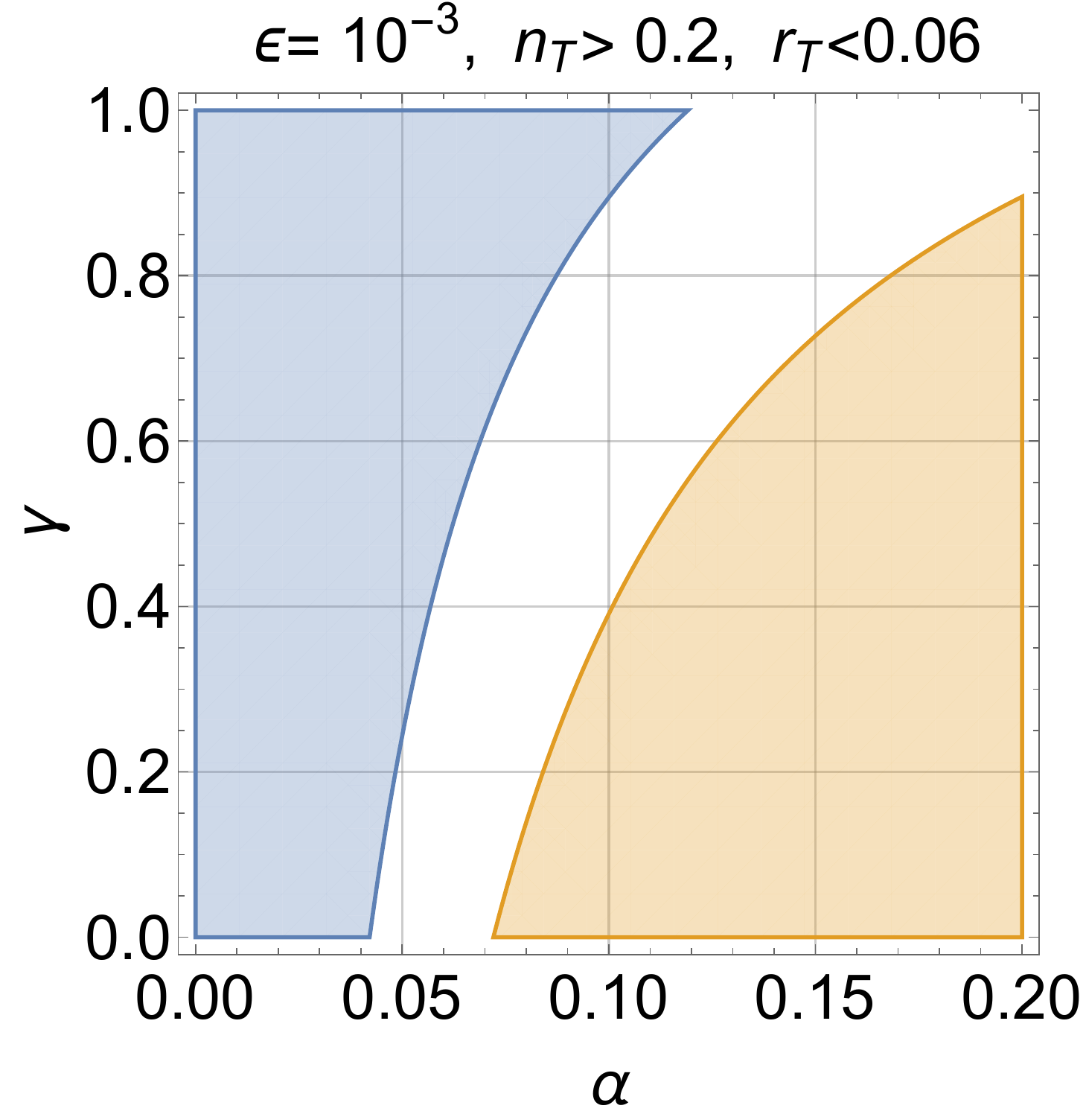}
\includegraphics[height=6.5cm]{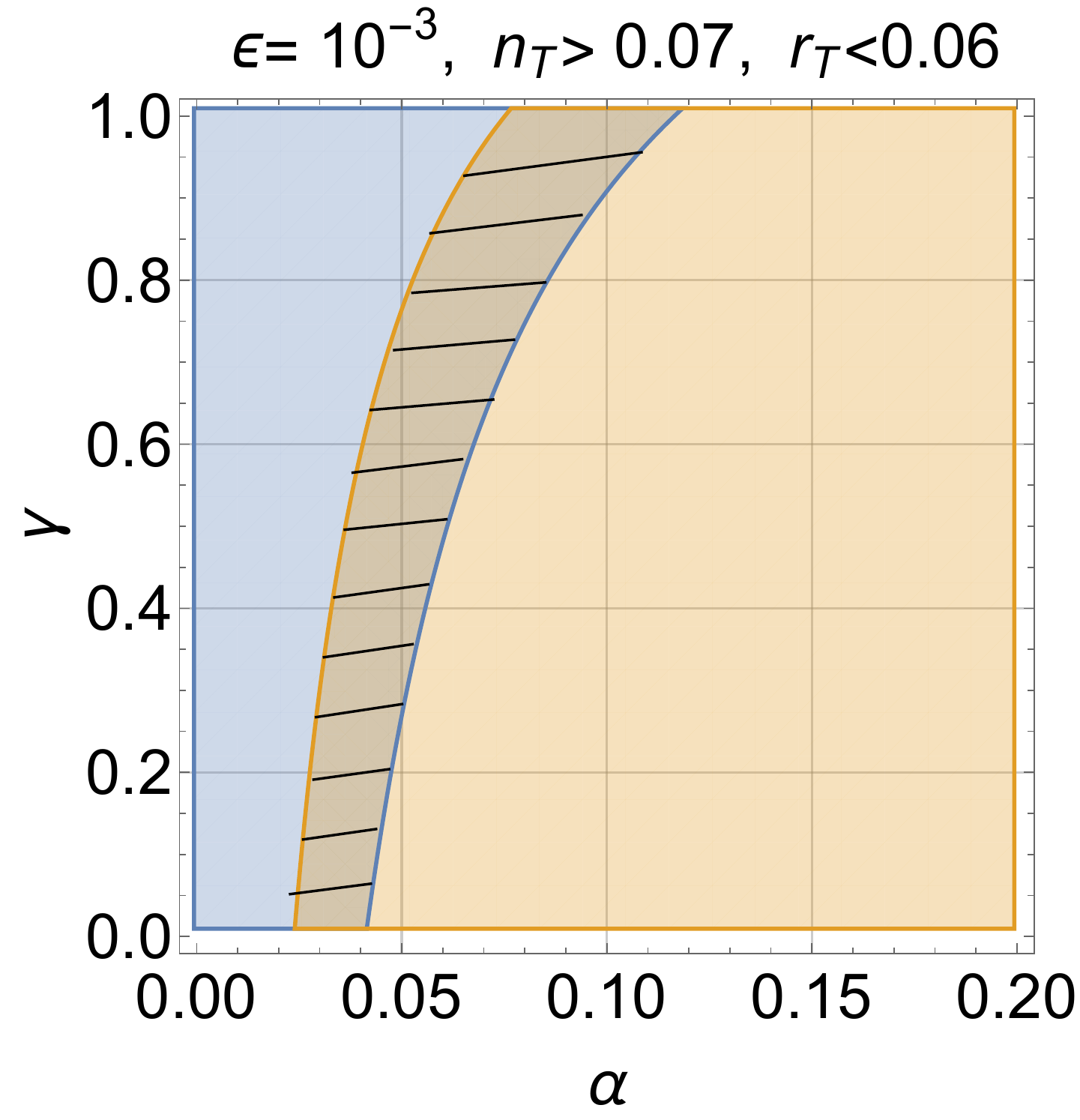}
\caption[a]{The constraints on $r_{T}$ and $n_{T}$ are translated in the $(\alpha,\,\gamma)$ plane for the values of $n_{T}$ 
compatible with an observable signal in the mHz band typical of space-borne interferometers.}
\label{FIGU5}      
\end{figure}

When $ N_{*} = {\mathcal O}(N_{t})$ the spectrum consists of only 
two branches: the standard infrared branch in the aHz region and 
an increasing branch extending up to the GHz. 
Since in this subsection we are discussing the case $N_{*} = {\mathcal O}(N_{t})$ 
the only two parameters are, in practice, $\alpha$ and $\gamma$.  
The problem is therefore to pin down the regions in the $(\alpha,\,\gamma)$ plane 
where the spectrum is blue, the signal is large and the constraints are not violated. 
The result of this analysis is illustrated in Figs. \ref{FIGU4} and \ref{FIGU5} where the parameters 
$N_{t}$ and $\epsilon$ have been selected as\footnote{Different values of these two fiducial parameters do not 
affect the conclusions of the present phenomenological discussion.}
 respectively, $70$ and $10^{-3}$. The conclusion suggested by Figs. \ref{FIGU4} and \ref{FIGU5} is that the 
 regions where the constraints are satisfied and the spectrum is sufficiently blue to guarantee a signal in the audio band do not overlap. 

Consider first the regions $r_{T} <0.06$ (from Eq. (\ref{MF15})) and $n_{T} >0.4$ 
(see Eq. (\ref{MF10a})): Fig. \ref{FIGU4} shows that these two corners of the parameter 
space are mutually exclusive. If we relax a bit the requirement on the spectral index 
and impose $n_{T} >0.3$ the two regions get a bit closer (see right plot in Fig. \ref{FIGU4}).
Whenever $N_{t} = {\mathcal O}(N_{*})$ the potential signals due to relic gravitons 
in the audio band are excluded and this result holds on spite of the value of $\gamma$. 
By looking at Fig. \ref{FIGU5} we see that the only appreciable overlap 
between the areas $n_{T} >0.07$ and $r_{T}<0.06$ is a 
tiny slice representing the allowed region of the parameter space.
By going back to Fig. \ref{FIGU3} we see that the spectral energy density corresponding to
 the allowed region of Fig. \ref{FIGU5}
is $h_{0}^2 \Omega_{\mathrm{gw}} = {\mathcal O}(10^{-15})$.  These values of the spectral energy density
will be hopefully probed by  Bbo/Decigo interferometer. We therefore conclude that in the case $N_{t}= N_{*}$ 
the variation of the refractive index does lead to a blue spectrum which is however only relevant in the 
 mHz band (but not in the audio band). In particular in the mHz band this signal can be 
 larger (by a couple of orders of magnitude) than the spectral energy density 
 coming from the conventional inflationary models.

 \subsection{Charting the parameter space in the $ N_{*} \ll  N_{t}$ case}
When $N_{*} < N_{t}$ (or even $N_{*}\ll N_{t}$) the spectral energy density consist 
of three branches: the usual infrared branch is supplemented by an intermediate 
regime (where the spectral slope is blue) terminating with a quasi-flat 
plateau at high frequencies. In the high-frequency region the spectrum 
is quasi-flat with a red slope determined by the value of $\epsilon$.
\begin{figure}[!ht]
\centering
\includegraphics[height=6.5cm]{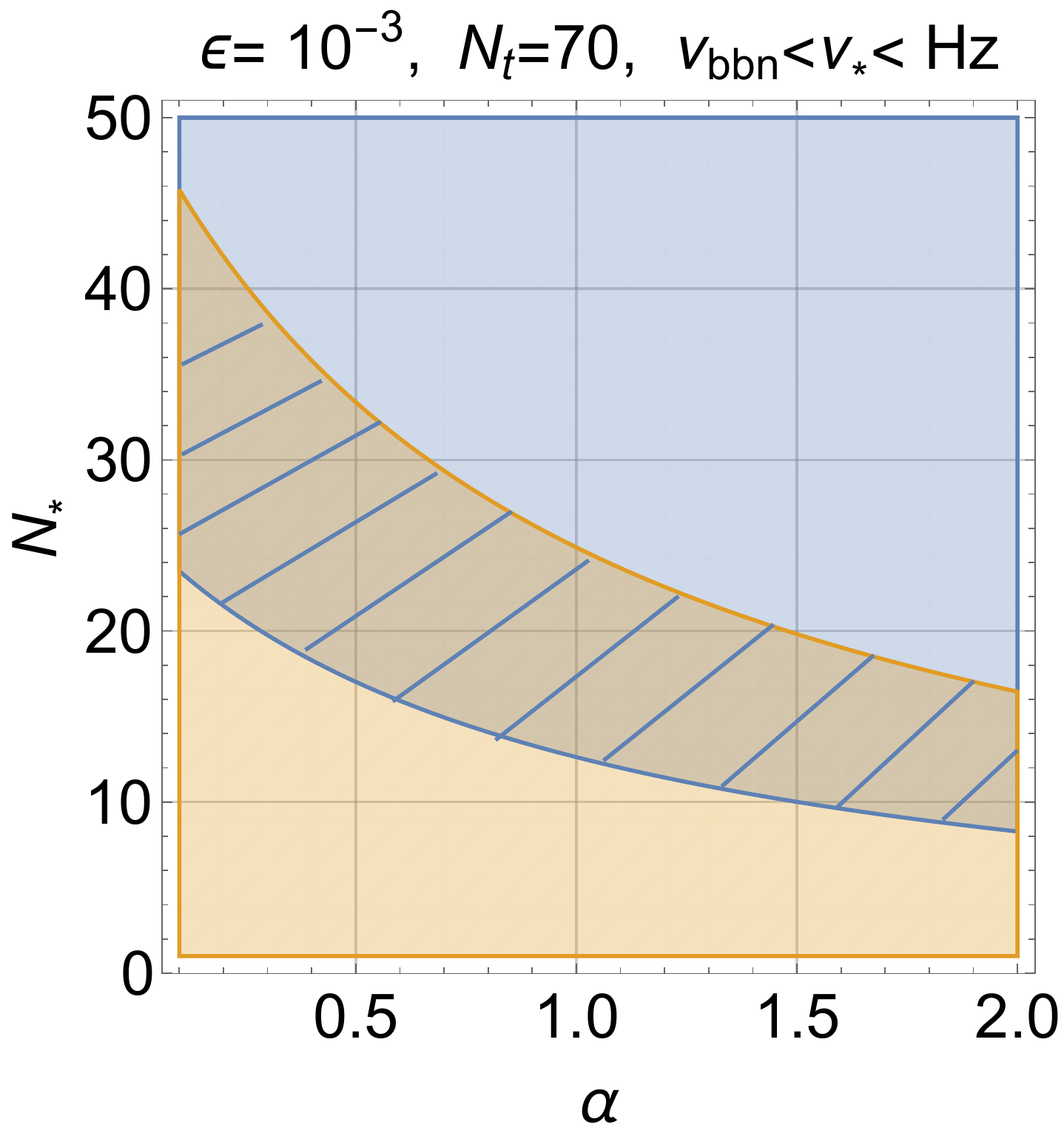}
\includegraphics[height=6.5cm]{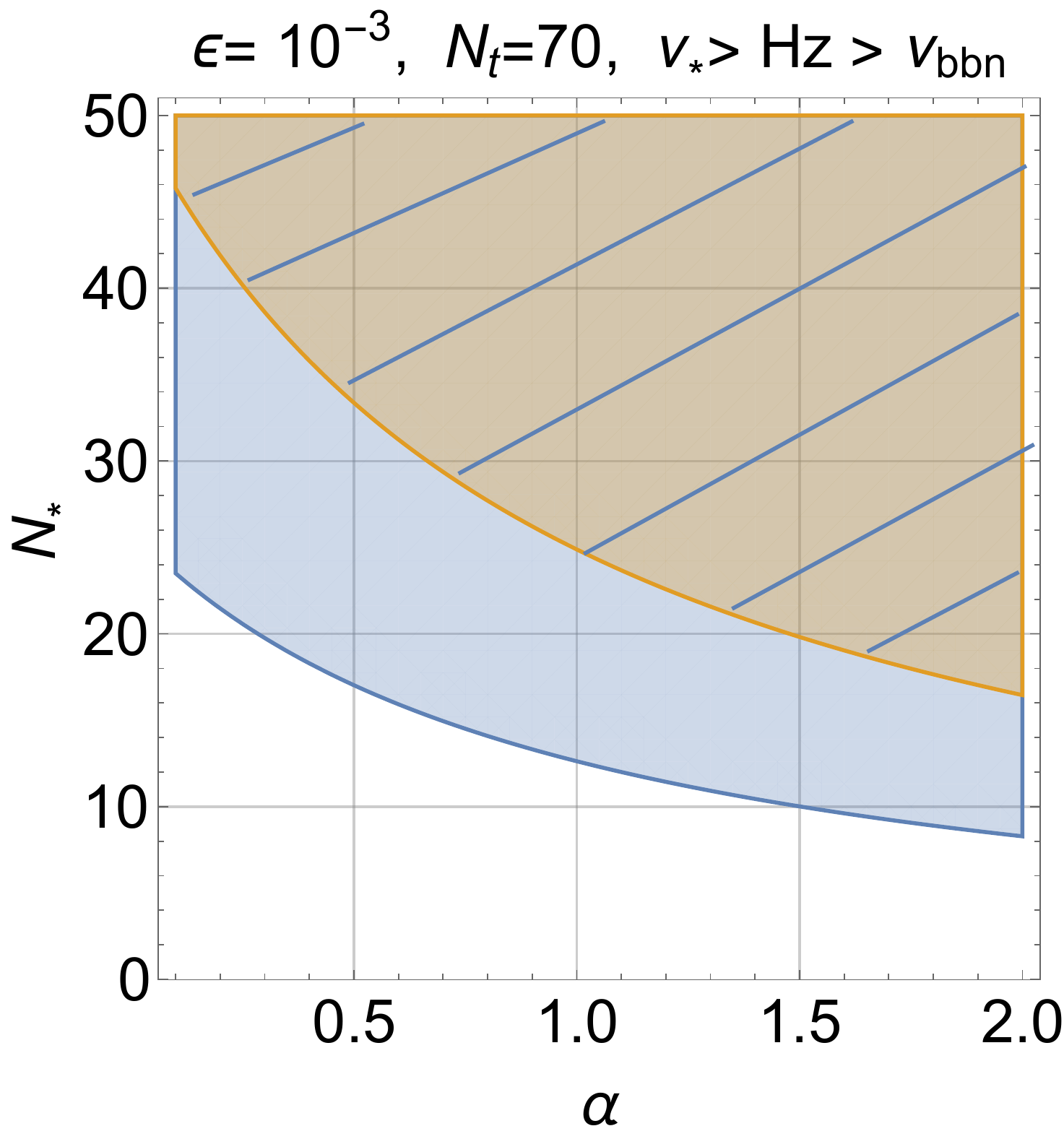}
\caption[a]{The regions $\nu_{\mathrm{bbn}} < \nu_{*} < \mathrm{Hz}$ and $\nu_{*}> \mathrm{Hz}$ are 
illustrated in the $(\alpha,\, N_{*})$ plane. The shaded area is determined by the overlap between the regions 
where the corresponding inequalities are satisfied. }
\label{FIGU6}      
\end{figure}
The frequency $\nu_{*}$ divides the intermediate branch of the spectrum 
from the high-frequency tail and it only depends on $N_{*}$, $N_{t}$ and $\alpha$.
There are two complementary physical possibilities.
 If $\nu_{\mathrm{bbn}} < \nu_{*} < \mathrm{Hz}$ the transition regime  takes place 
between the frequency of the big-bang nucleosythesis and the mHz 
band where space-borne interferometers will be operating; this region 
correspond to the shaded area appearing in the left plot of 
Fig. \ref{FIGU6}. In the complementary regime we have instead that 
$\nu_{*} > \mathrm{Hz}$ and the 
break of the spectrum is in the audio range. This region corresponds to the 
shaded area appearing in the right plot of Fig. \ref{FIGU6}.
As long as $25 < N_{*} < 50$ and $0<\alpha < 2$,
the results of Fig. \ref{FIGU6} confirm that 
$\nu_{\mathrm{bbn}} < \nu_{*} < \mathrm{Hz}$, as expected 
from qualitative arguments. For the same range of $\alpha$ the condition $\nu_{*}> \mathrm{Hz}$
will be verified for a comparatively larger $N_{*}$.

\begin{figure}[!ht]
\centering
\includegraphics[height=6.5cm]{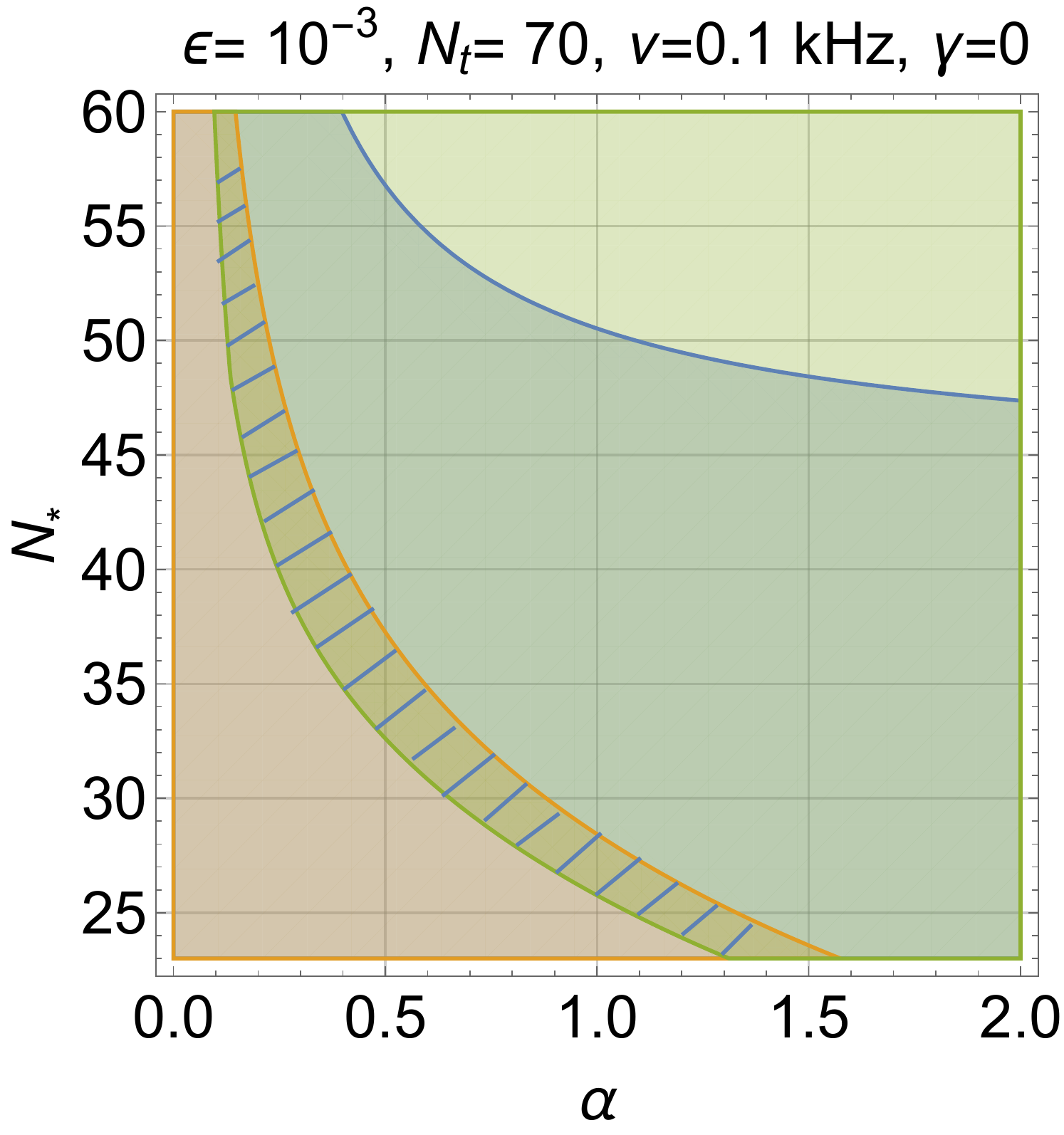}
\includegraphics[height=6.5cm]{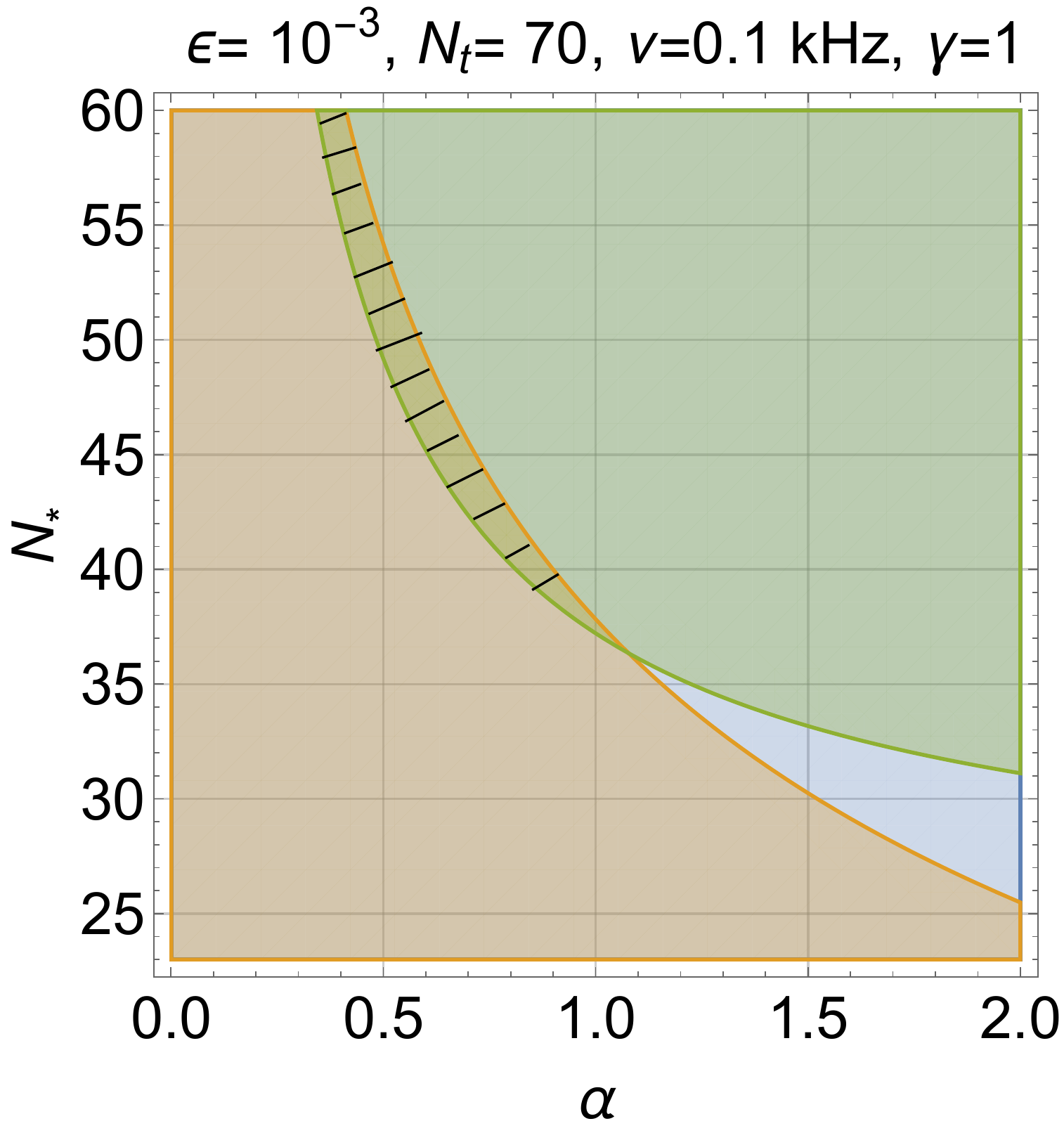}
\caption[a]{We illustrate the regions where the relevant phenomenological bounds are satisfied and the spectral energy density 
(in critical units) exceeds $10^{-11}$ at the centre of the audio band in the cases $\gamma =0$ (left plot) and $\gamma =1$ 
(right plot). The dashed areas denote the regions  where the constraints are simultaneously 
satisfied and $h_{0}^2 \Omega_{\mathrm{gw}}(\nu_{LV},\tau_{0}) > 10^{-11}$. }
\label{FIGU7}      
\end{figure}

We are now in conditions of charting the parameter space in the $(\alpha, \, N_{*})$ 
plane by requiring that the constraints of Eqs. (\ref{CC1})
Eqs. (\ref{CC2}) and (\ref{CC1a}) are simultaneously satisfied.
In Fig. \ref{FIGU7} we enforced all the relevant bounds 
and also required\footnote{The expression of the signal-to-noise ratio in the context of optimal processing 
depends on a number of factors including the spectral slope and the overlap reduction. The requirements 
of this section on the target signals must therefore be considered as necessary but, strictly speaking, not sufficient.}
\begin{equation}
h_{0}^2 \Omega_{\mathrm{gw}}(\nu_{LV},\tau_{0}) > 10^{-11},\qquad  \nu_{LV} =0.1 \, \mathrm{ kHz},
\label{audio}
\end{equation}
where $\nu_{LV}$ denotes the Ligo/Virgo frequency falling within the audio band.
 While the variation of $\gamma$ can also be considered 
continuous in the purely heuristic perspective invoked after Eq. (\ref{fifth}),
we shall just illustrate the cases $\gamma= 0$ and $\gamma=1$ 
corresponding to the two equivalent parametrizations of the action 
discussed in sections \ref{sec2} and \ref{sec3}.
In this connection we can remark that in the case $\gamma=0$ (left plot in Fig. \ref{FIGU7}) 
the allowed region is larger than in the case $\gamma =1$ (right plot in Fig. \ref{FIGU7}). 
Note that in Fig. \ref{FIGU7} the frequency $\nu_{a}=0.1 \mathrm{kHz}$ has been 
chosen as representative of the typical signal in the audio-band where the sensitivity 
of wide-band detectors is approximatively larger. A different choice in the range 
$\mathrm{Hz} < \nu_{a} < \mathrm{kHz}$ will lead to similar conclusions\footnote{It is 
interesting to remark that the regions $\alpha < 0.1$ and $N_{*} \to N_{t}$ (i.e. on the left of both 
plots) are excluded and this is consistent with the results obtained above in this section 
for the case $N_{*} = {\mathcal O}(N_{t})$.}.
 
\begin{figure}[!ht]
\centering
\includegraphics[height=6cm]{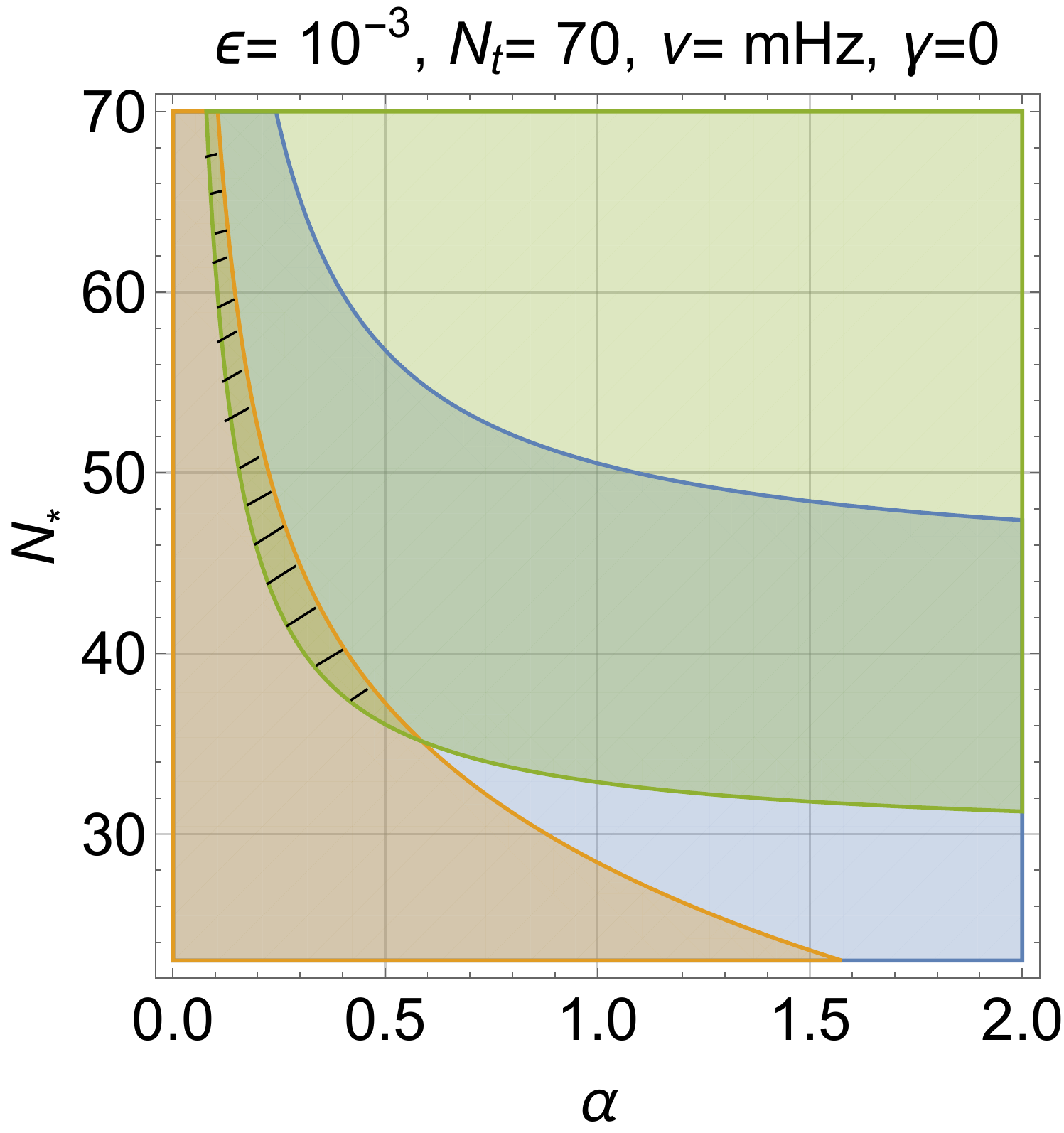}
\includegraphics[height=6cm]{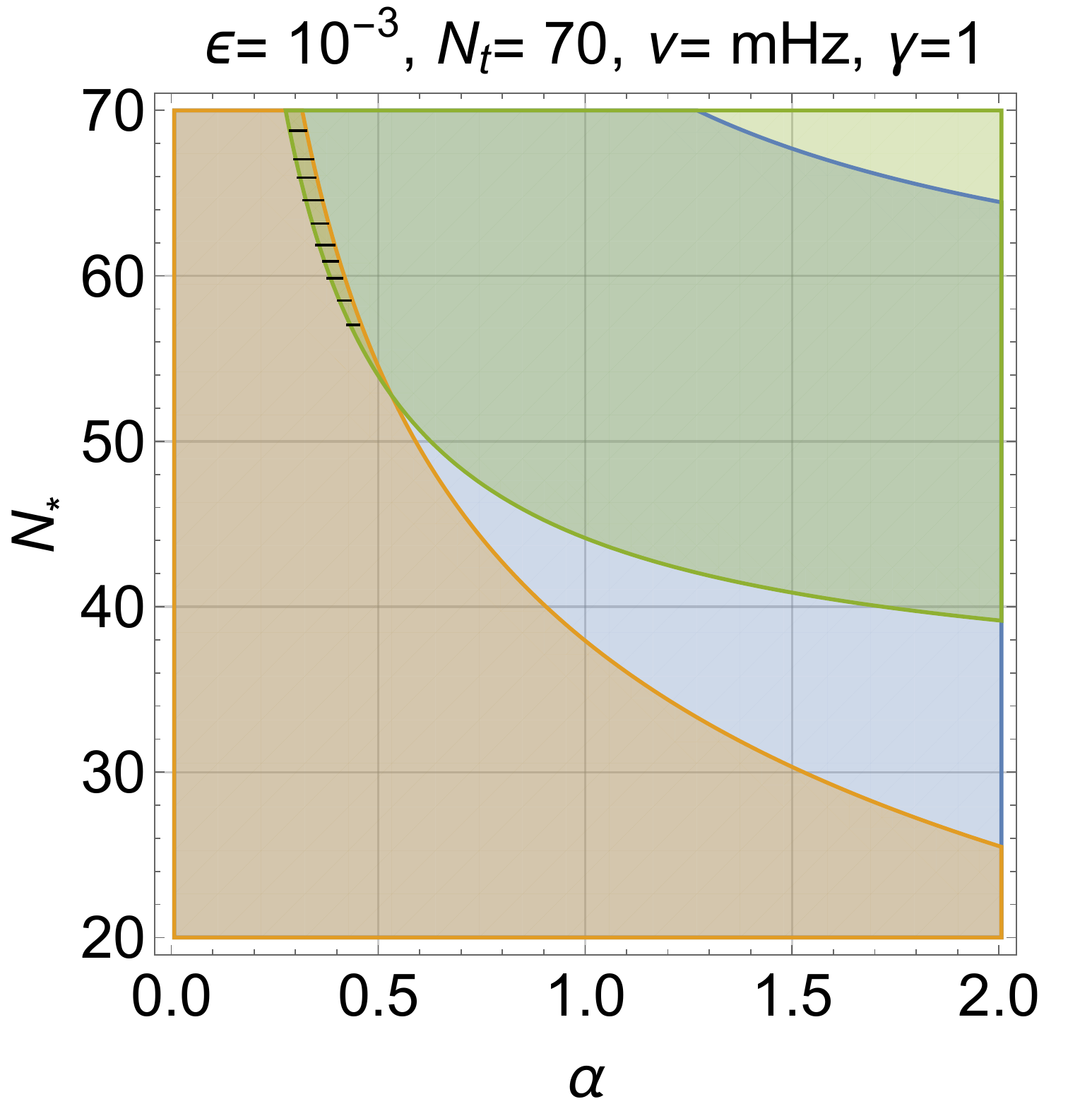}
\caption[a]{We illustrate the regions where the relevant phenomenological bounds are satisfied and the spectral energy density 
(in critical units) exceeds $10^{-12}$ at the centre of the mHz band in the cases $\gamma =0$ (left plot) and $\gamma =1$ 
(right plot). The dashed areas denote the regions of the parameter space where the three constraints are simultaneously 
satisfied. Figures \ref{FIGU7} and \ref{FIGU8} pin down the two complementary portions 
of the parameter space: by comparison we can see that the dashed areas of Figs. \ref{FIGU7} and \ref{FIGU8} do overlap.}
\label{FIGU8}      
\end{figure}

In Fig. \ref{FIGU8} the same analysis of Fig. \ref{FIGU7} is extended to
mHz band. The dashed areas in Fig. \ref{FIGU8} have the same meaning 
of those depicted in Fig. \ref{FIGU7} with the difference that we are 
now imposing a more stringent requirement on the spectral energy density namely:
\begin{equation}
h_{0}^{2} \Omega_{\mathrm{gw}}(\nu_{m}, \tau_{0}) > 10^{-12}, \qquad \nu_{m}= 10^{-3}\, \mathrm{Hz}.
\label{milli}
\end{equation}
The comparison of Figs. \ref{FIGU7} and \ref{FIGU8} demonstrates that there exist 
a tiny overlap between the shaded areas appearing in the two figures.
The overlap between the dashed areas of  Figs. \ref{FIGU7} and \ref{FIGU8} 
defines, for each value of $\gamma$,  a sweet spot in the $(\alpha, \, N_{*})$ plane where
 the signal is be potentially visible both in the audio band and in the mHz band. 

\begin{figure}[!ht]
\centering
\includegraphics[height=6cm]{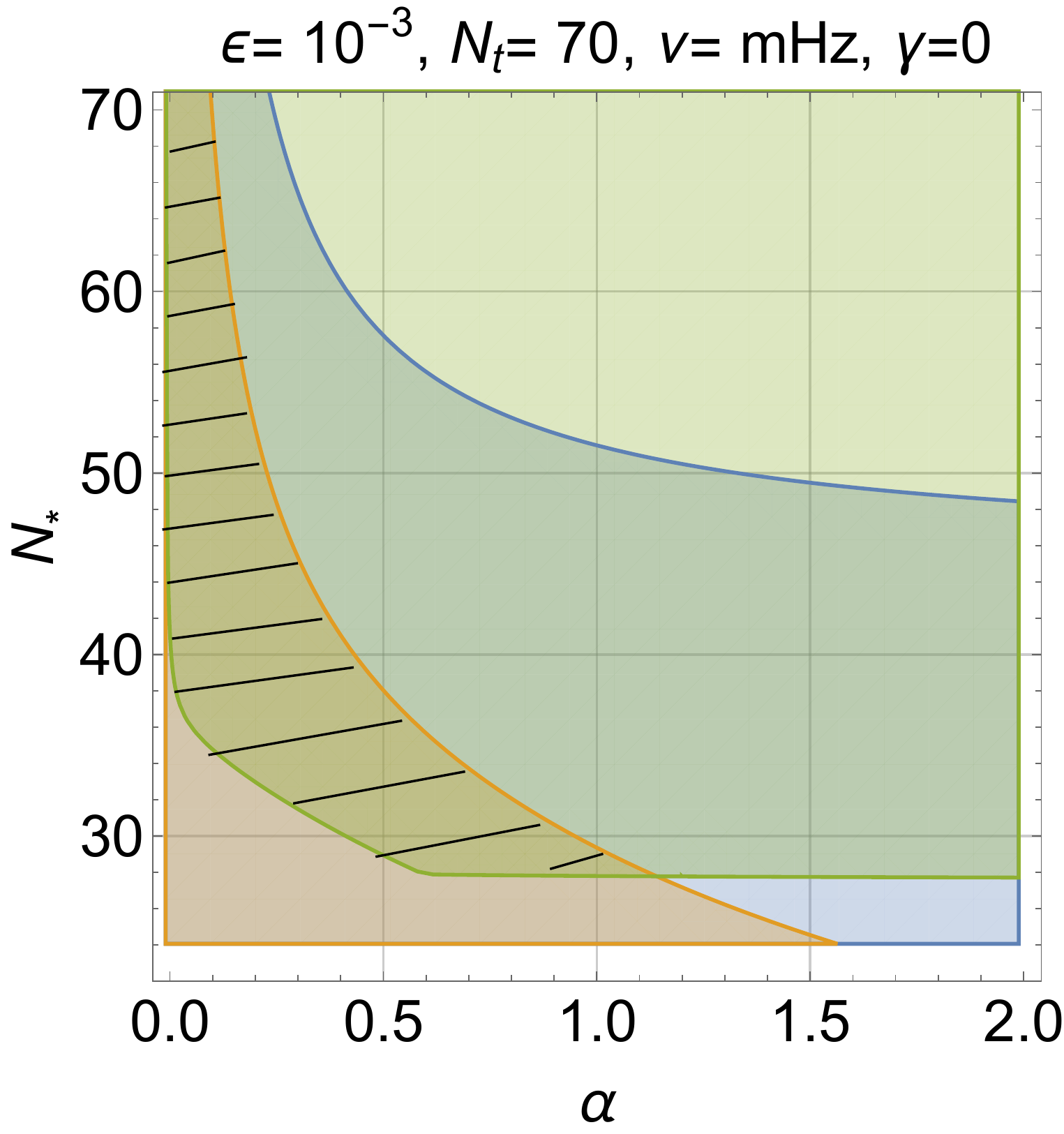}
\includegraphics[height=6cm]{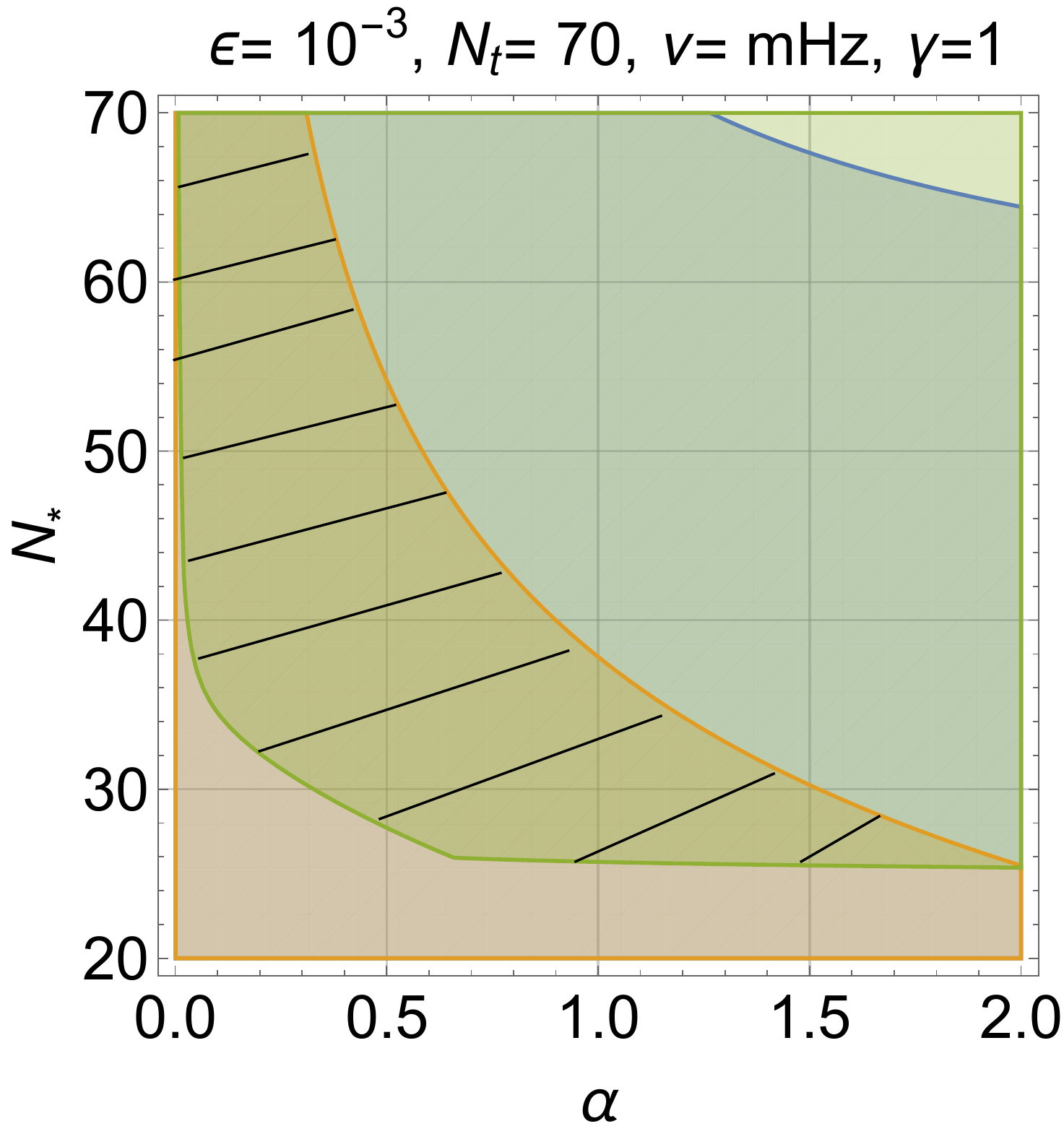}
\caption[a]{We illustrate the regions where the relevant phenomenological bounds are satisfied and the spectral energy density 
is critical units exceeds $10^{-16}$ at the centre of the mHz band in the cases $\gamma =0$ (left plot) and $\gamma =1$ 
(right plot). The dashed areas denote, as usual, the regions of the parameter space where the three constraints are simultaneously 
satisfied.}
\label{FIGU9}      
\end{figure}

If we would require a larger sensitivity in the mHz band (such as the the ones often quoted for the Bbo or Decigo projects) 
we would be led to demand, for instance, 
\begin{equation}
h_{0}^{2} \Omega_{\mathrm{gw}}(\nu_{m}, \tau_{0}) > 10^{-16}, \qquad \nu_{m}= 10^{-3}\, \mathrm{Hz}.
\label{milli2}
\end{equation}
 In Fig. \ref{FIGU9} the Bbo/Decigo target sensitivities are illustrated in the $(\alpha,\,N_{*})$ plane. As it 
is clearly visible from the direct comparison of Figs. \ref{FIGU8} and \ref{FIGU9} the dashed 
areas get larger. Still the allowed region in the  $\gamma = 1$ case is always larger than for $\gamma =0$. 
By comparing Figs. \ref{FIGU7}, \ref{FIGU8} and \ref{FIGU9}  in a unified perspective, the sweet spots now get
larger if we demand that space-borne interferometers will be able to probe 
values of the spectral energy density as low as the ones of Eq. (\ref{milli2}). 
In particular the dashed area of Fig. \ref{FIGU7} is now all contained within the dashed 
area of Fig. \ref{FIGU9}: this means that by remaining in the dashed slice of Fig. \ref{FIGU7}
the resulting signal will be, a fortiori, visible both in the audio band and in the mHz band.

Even if the target sensitivities of Eq. (\ref{milli2}) look by definition quite optimistic they might not be totally 
insane if we bluntly assume that the minimal detectable chirp (or strain) amplitudes are the same in the 
mHz and in the audio band. The argument goes in short as follows.
A given chirp (or strain amplitude) of the signal in the audio band (say for $\nu=\nu_{LV} = {\mathcal O}(0.1) \mathrm{kHz}$) 
probes a spectral energy density which is comparatively larger than the one achievable, with the same 
sensitivity, in the mHz band. To appreciate this we can consider, for instance, Eq. (\ref{ref1}) and suppose that the spectral 
energy density of the signal will be, in critical units, $h_{0}^2 \Omega_{\mathrm{gw}}(\nu_{a}, \tau_{0}) = {\mathcal O}(10^{-10})$.
The required sensitivity in terms of $h_{c}$ will therefore be of the order of $10^{-25}$. 
Let us now pretend that the achievable sensitivity (in terms of the chirp amplitude and in the mHz band) is the same as in the audio band, i.e. 
$h_{c}= {\mathcal O}(10^{-25})$. Since the frequency in the mHz band is ${\mathcal O}(10^{-3})$ Hz, Eq. (\ref{ref1}) can be 
inverted in order to obtain the minimal spectral amplitude detectable in the mHz band; the result will be $h_{0}^2 \Omega_{\mathrm{gw}}(\nu_{a}, \tau_{0}) = {\mathcal O}(10^{-15})$. 

We shall now select 
the values of $\alpha$ and $N_{*}$ within the dashed areas of Figs. \ref{FIGU7}, \ref{FIGU8} and \ref{FIGU9} and
present some interesting example.
\begin{figure}[!ht]
\centering
\includegraphics[height=6.5cm]{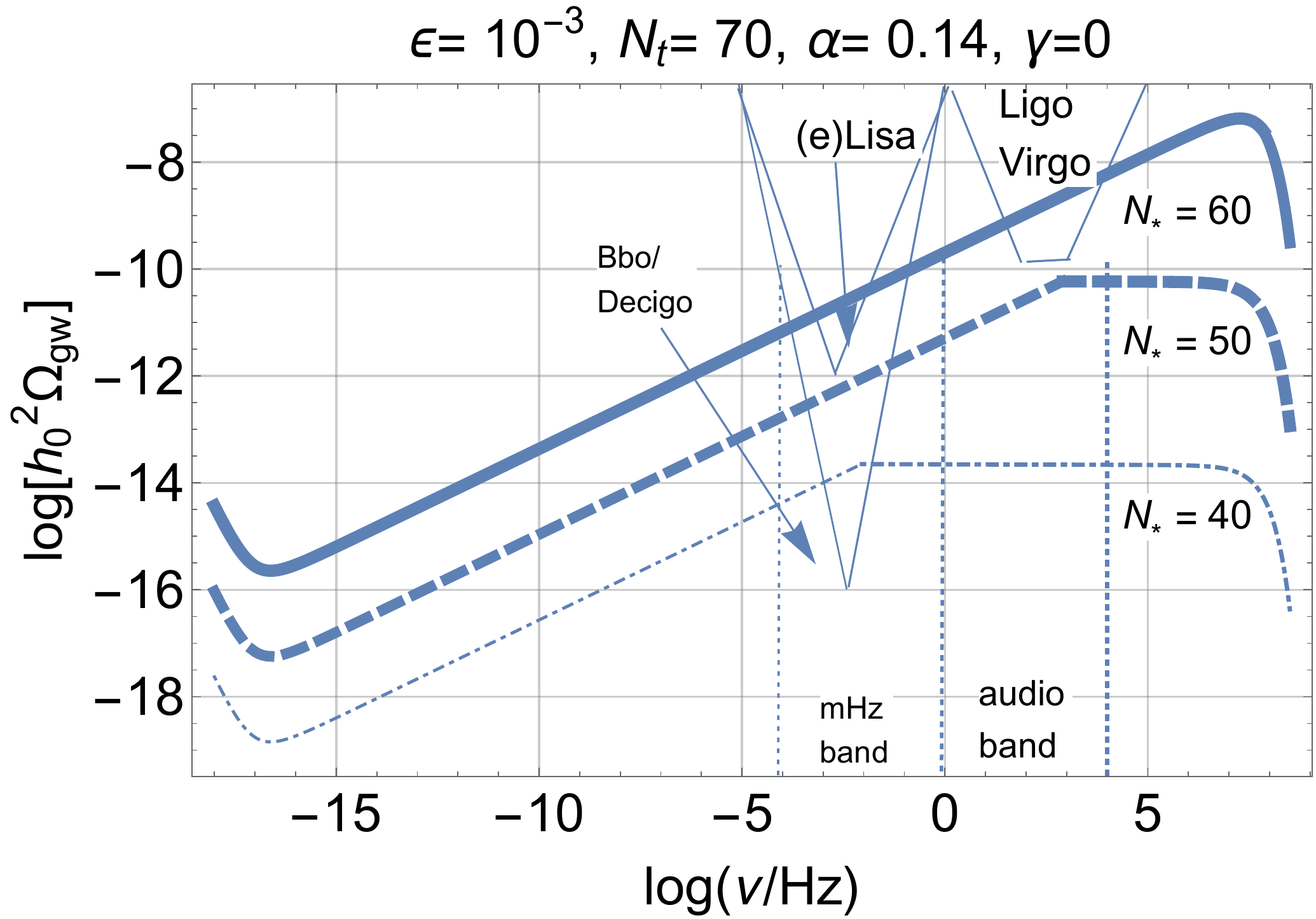}
\caption[a]{The spectral energy density is illustrated for various choices of $N_{*}$ with  $N_{*} < N_{t}$. The other 
values have been selected within the allowed regions 
of the parameter space and they are $\alpha = 0.14$ and $\gamma =0$. }
\label{FIGU10}      
\end{figure}
In Fig. \ref{FIGU10} we illustrate the spectral energy density by fixing $\alpha = 0.14$ (for $\gamma=0$); 
similarly in  Fig. \ref{FIGU11} we demand $\alpha =0.42$ (for $\gamma =1$). Both in Figs. \ref{FIGU10} and \ref{FIGU11}
we choose different values of $N_{*}$ (i.e. $N_{*}= 40,\, 50,\, 60$). As $N_{*} \to N_{t}$ the 
results discussed at the beginning of this section are recovered: the high-frequency plateau gets 
narrower and asymptotically disappears. Conversely, when the value of $N_{*}$ decreases, $\nu_{*}$ also diminishes and the quasi-flat branch 
of the spectrum gets wider, as it can be seen from Figs. \ref{FIGU10} and \ref{FIGU11}.

Let us now consider more closely the values of $\alpha$ and $\gamma$. 
In the left plot of Fig. \ref{FIGU9} (i.e. $\gamma =0$) the value of $\alpha =0.14$ is within the shaded region and the same is true for $\alpha =0.42$ in the right plot (i.e. $\gamma = 1$) of the same figure. The value of $\alpha$ in Fig. \ref{FIGU10} is three times smaller than the one of Fig. \ref{FIGU11} and this 
occurrence has a simple explanation\footnote{The spectral index $n_{T}$, according to Eqs. (\ref{MF11})--(\ref{MF12}), is roughly 
given by $3\alpha/(\alpha +1) + {\mathcal O}(\epsilon)$ in the case $\gamma =0$ while it is 
$\alpha/(\alpha +1) + {\mathcal O}(\epsilon)$ for $\gamma =1$.  For the same spectral index $\overline{n}_{T}$ 
we have that, roughly, $\alpha^{(\gamma=0)} = \overline{n}_{T}/3 \simeq \alpha^{(\gamma=1)}/3$. 
This is the reason why the slopes in the intermediate 
branch are approximately the same between Figs \ref{FIGU10} and \ref{FIGU11}.} which is directly related to Eqs.  (\ref{MF11})--(\ref{MF12}).
\begin{figure}[!ht]
\centering
\includegraphics[height=6.5cm]{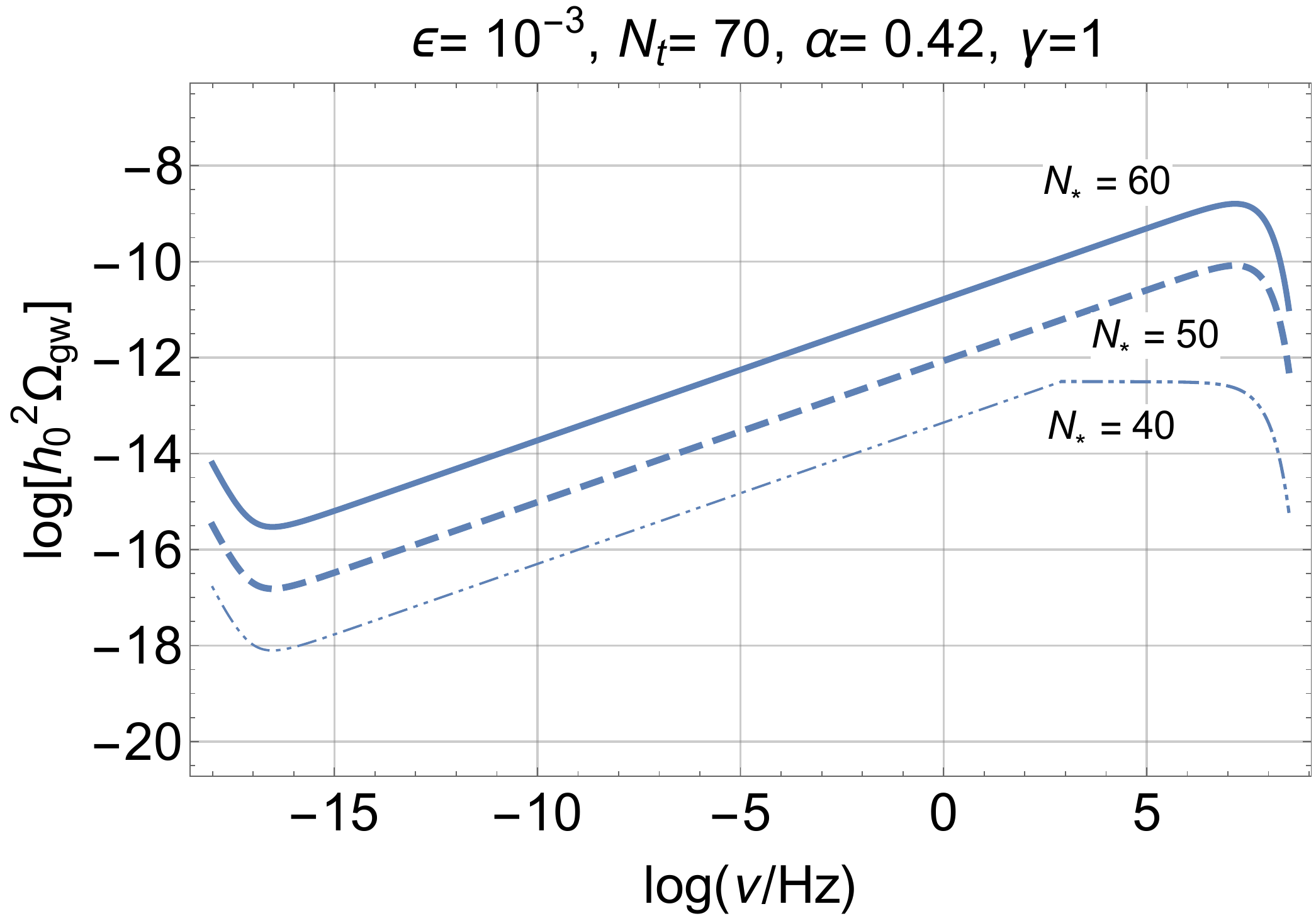}
\caption[a]{The spectral energy density is illustrated for  $N_{*} < N_{t}$; to compare this plot with the one of Fig. \ref{FIGU10} we stress that
  $\alpha = 0.42$ and $\gamma = 1$. }
\label{FIGU11}      
\end{figure}
By looking at Figs. \ref{FIGU10} and \ref{FIGU11} we see that for a value of $\alpha$  
leading to the same $n_{T}$ in the cases $\gamma =0$ and $\gamma =1$, 
the spectral energy density in the case $\gamma=1$ is smaller than in the case $\gamma =0$ 
for the same choice of $N_{*}$.

All in all, when $N_{*} < N_{t}$ the spectral 
energy density develops two branches for $\nu> \nu_{\mathrm{eq}}$. The intermediate branch corresponds to those 
modes leaving the Hubble radius during inflation when the refractive index is still dynamical. These modes 
will then lead to an increasing spectral slope for $ \nu_{\mathrm{eq}} < \nu< \nu_{*}$. 
The second branch of the spectrum corresponds to those modes that left the Hubble radius 
when the refractive index already ceased to be dynamical: in this branch the spectrum decreases 
as $\nu^{- 2 \epsilon}$ for $\nu_{*} < \nu < \nu_{\mathrm{max}}$.In Fig. \ref{FIGU10} the audio branch of the spectrum and the mHz branch 
have been illustrated, respectively,  with trapeziodal and triangular shapes encompassing the respective 
frequency ranges. The heights of the two triangles roughly matches the hoped sensitivities 
of (e)Lisa and Bbbo/Decigo. Similarly the height of the trapezoid in the audio branch 
approximately correspond to the sensitivities of terrestrial wide-band detectors (in their advanced version). 
Not all the curves of Figs. \ref{FIGU10} and \ref{FIGU11} match the required sensitivities 
of future detectors. 

\begin{figure}[!ht]
\centering
\includegraphics[height=5.5cm]{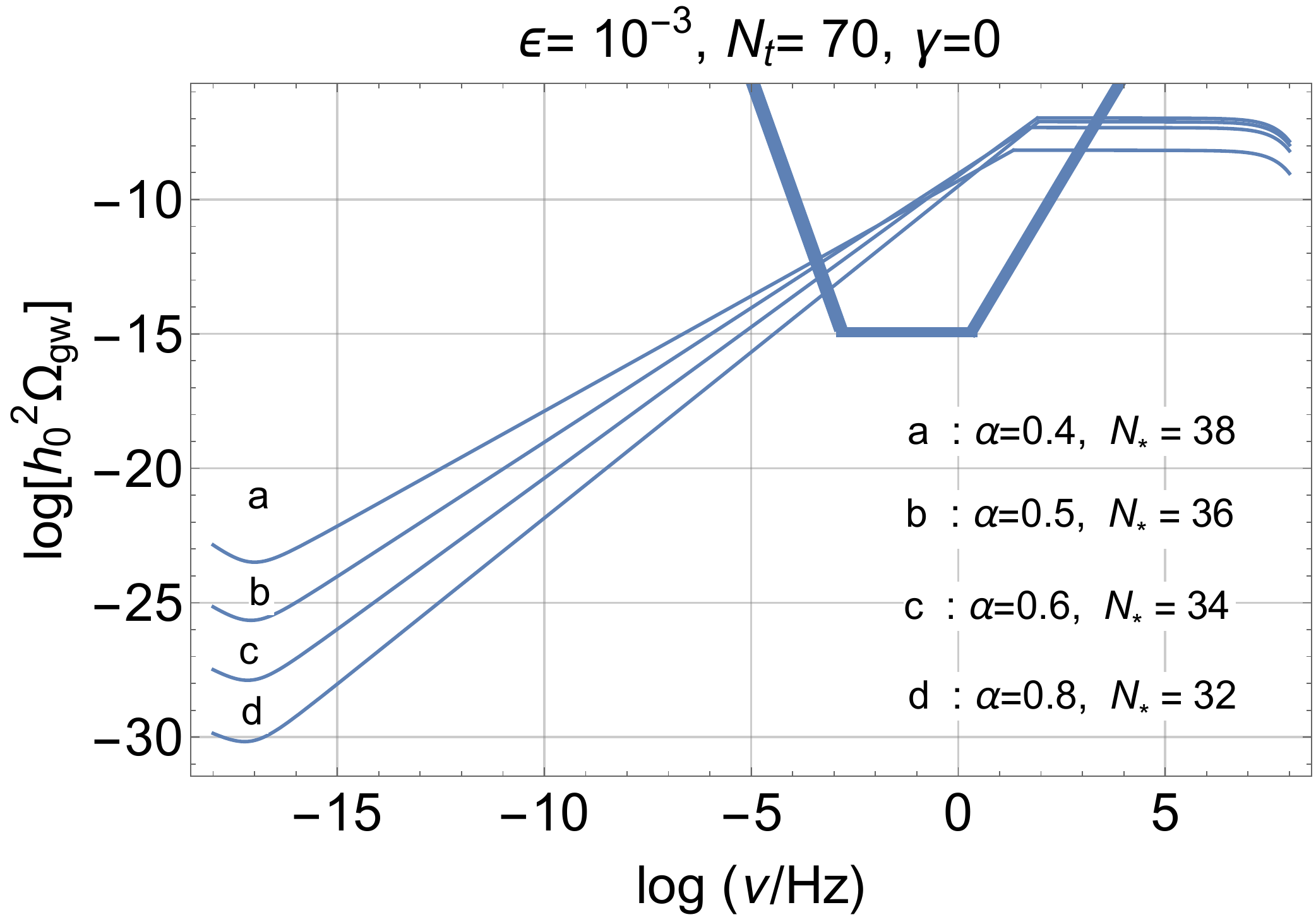}
\includegraphics[height=5.5cm]{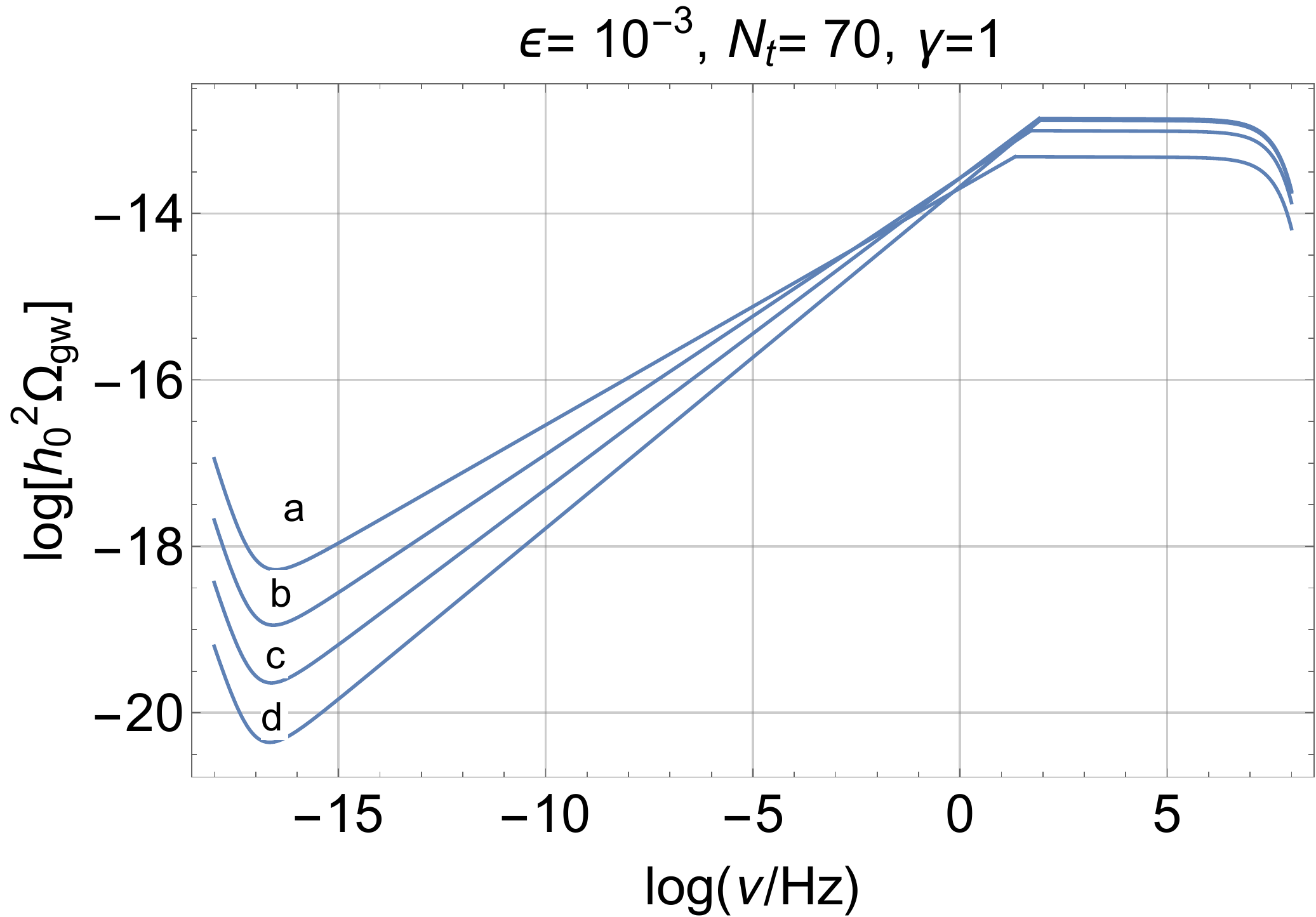}
\caption[a]{The spectral energy density in critical units for a peculiar choice 
of the parameters selected within the shaded regions in Fig. \ref{FIGU9} (plot at the left). }
\label{FIGU12}      
\end{figure}

In the $(\alpha,\,N_{*})$ plane there exist regions that 
are potentially visible both in the audio band and in the mHz bands. These regions 
lie within the shaded areas of Figs. \ref{FIGU7} and \ref{FIGU9} and are illustrated 
in Fig. \ref{FIGU12} in the case $\gamma =0$.
The trapezoidal shape of Fig. \ref{FIGU12} generously encompasses the 
best sensitivities in the spectral energy density for terrestrial and space-borne
detectors in the audio and mHz bands. The parameters illustrated in Fig. \ref{FIGU12} 
belong to the same fine-tuning line: if the parameters lie along that line in the $(\alpha, \, N_{*})$ 
plane we can be reasonably sure that the corresponding spectral energy density in 
critical units will be potentially visible both by terrestrial and by space-borne 
detectors. The  parameters leading to a comparatively large spectral energy density
in the case $\gamma =0$ correspond to a much smaller value of $h_{0}^2 \Omega_{\mathrm{gw}}$ in the case 
$\gamma=1$. This conclusion is clear by comparing the two plots of Fig. \ref{FIGU12}:
the two plots share exactly the same parameters but correspond to different values of $\gamma$. 
As explained above in connection with Figs. \ref{FIGU10} and \ref{FIGU11} this 
feature has a simple explanation since different values of $\gamma\neq 0$ simply shift 
the tensor spectral index of the intermediate branch in comparison with the case $\gamma =0$.

\renewcommand{\theequation}{5.\arabic{equation}}
\setcounter{equation}{0}
\section{Concluding remarks}
\label{sec5}
We showed that a rather generic scenario for the evolution 
of the refractive index during the early stages 
of a conventional inflationary phase leads 
to power spectrum of relic gravitons naturally 
tilted towards frequencies. The spectral slope 
is determined, in this context, by the competition 
between the rate of variation of the refractive index 
 and the expansion rate of the quasi-de Sitter stage.
We argued that it is always possible to redefine the time 
coordinate and to compute the evolution 
of the field operators in terms of an effective 
horizon incorporating the dynamical effects due to the 
variation of the refractive index. 

The slope of the tensor spectrum differs for the modes 
crossing the effective horizon in the subsequent phase 
where the  gravitons propagate again at the 
speed of light and the refractive index goes back to $1$.
In this branch the spectral slope is quasi-flat 
but it decreases at a rate that depends on the slow-roll
parameter. If the evolution of the refractive index stops 
at the end of inflation, the spectral energy density   
consists of the usual low-frequency branch (between 
few aHz and $0.1$ fHz) supplemented 
by a growing branch extending up to the MHz or even GHz regions.
Conversely if the evolution of the  refractive index terminates before the 
end of inflation the previous branches are supplemented 
by a quasi-flat plateau at high frequencies. The frequency range 
of the plateau depends on the rate of variation of the refractive index 
(in units of the expansion rate) and on the critical number of efolds 
during which the refractive index evolves. After imposing the 
usual phenomenological constraints, it turns out that 
the plateau may extend  from the mHz band 
to the GHz band. In a sweet spot of the parameter space all the 
known bounds  are satisfied and the 
resulting signal could be detectable, in principle, either 
by terrestrial interferometers (in their advanced and enhanced 
configuration) or by space-borne detectors. 

\section*{Acknowledgements}

It is a pleasure to thank F. Fidecaro for interesting discussions.

\newpage
\begin{appendix}
\renewcommand{\theequation}{A.\arabic{equation}}
\setcounter{equation}{0}
\section{Tensor, scalars and their frame-invariance}
\label{APPA}
The two polarizations of the gravitational wave are defined as:
 \begin{equation}
 e_{ij}^{(\oplus)}(\hat{k}) = (\hat{m}_{i} \hat{m}_{j} - \hat{q}_{i} \hat{q}_{j}), \qquad 
 e_{ij}^{(\otimes)}(\hat{k}) = (\hat{m}_{i} \hat{q}_{j} + \hat{q}_{i} \hat{m}_{j}),
 \label{ST0}
 \end{equation}
where  $\hat{k}_{i} = k_{i}/|\vec{k}|$,  $\hat{m}_{i} = m_{i}/|\vec{m}|$ and $\hat{q} =q_{i}/|\vec{q}|$ are three mutually 
orthogonal directions  and $\hat{k}$. It follows directly from Eq. (\ref{ST0}) that $e_{ij}^{(\lambda)}\,e_{ij}^{(\lambda')} = 2 \delta_{\lambda\lambda'}$ while the sum over the polarizations gives:
\begin{equation}
\sum_{\lambda} e_{ij}^{(\lambda)}(\hat{k}) \, e_{m n}^{(\lambda)}(\hat{k}) = \biggl[p_{m i}(\hat{k}) p_{n j}(\hat{k}) + p_{m j}(\hat{k}) p_{n i}(\hat{k}) - p_{i j}(\hat{k}) p_{m n}(\hat{k}) \biggr];
\label{ST0B} 
\end{equation}
where $p_{ij}(\hat{k}) = (\delta_{i j} - \hat{k}_{i} \hat{k}_{j})$. Defining the Fourier transform of $\hat{h}_{ij}(\vec{x},\eta)$ as 
\begin{equation}
\hat{h}_{ij}(\vec{x},\eta) = \frac{1}{(2\pi)^{3/2}}\, \sum_{\lambda}  \, \int d^{3} k \, \hat{h}_{ij}(\vec{k},\eta)\,\, e^{- i \vec{k}\cdot\vec{x}},
\label{St0C}
\end{equation}
we  also have that:
\begin{equation}
\hat{h}_{ij}(\vec{p}, \eta) = \sum_{\lambda} \biggl[ e_{ij}^{(\lambda)}(\hat{p}) F_{k,\, \lambda}(\eta) \hat{a}_{\vec{p}\, \lambda} +
e_{ij}^{(\lambda)}(-\hat{p})  F_{k,\, \lambda}^{*}(\eta)  \hat{a}_{-\vec{p}\, \lambda}^{\dagger} \biggr], 
\label{TTT1}
\end{equation}
The tensor power spectra ${\mathcal P}_{T}(k,\eta)$ and ${\mathcal Q}_{T}(k,\eta)$ determine the two-point function at equal times:
\begin{equation}
\langle \hat{h}_{ij}(\vec{k},\eta) \, \hat{h}_{mn}(\vec{p},\eta) \rangle = \frac{2\pi^2}{k^3} {\mathcal P}_{T}(k,\eta) \, {\mathcal S}_{ijmn}(\hat{k}) \delta^{(3)}(\vec{k} +\vec{p}),
\label{ST1}
\end{equation}
where ${\mathcal P}_{T}(k,\eta)$ is the power spectrum in the $\eta$ parametrization while: 
\begin{equation}
{\mathcal S}_{ijmn}(\hat{k}) = \sum_{\lambda} e_{ij}^{(\lambda)}(\hat{k}) \, e_{m n}^{(\lambda)}(\hat{k})/4 = 
\frac{1}{4} \biggl[ p_{mi}(\hat{k}) p_{nj}(\hat{k}) + p_{mj}(\hat{k}) p_{ni}(\hat{k}) - p_{ij}(\hat{k}) p_{m n}(\hat{k}) \biggr].
\label{ST1a}
\end{equation}
Equation (\ref{ST1}) follows the same conventions used when deriving the spectrum of curvature perturbations  on comoving orthogonal hypersurfaces (customarily denoted by ${\mathcal R}(\vec{x},\tau)$) 
\begin{equation}
\langle {\mathcal R}(\vec{k},\tau) \, {\mathcal R}(\vec{p},\tau) \rangle = \frac{2\pi^2}{k^3} {\mathcal P}_{{\mathcal R}}(k,\tau)  \delta^{(3)}(\vec{k} +\vec{p}),
\label{ST1C}
\end{equation}
which is the quantity employed to set the initial conditions for the evolution of the temperature and polarization anisotropies of the Cosmic Microwave Background. According to the standard convention,
the scalar power spectrum is assigned as:
\begin{equation}
{\mathcal P}_{{\mathcal R}}(k)= {\mathcal A}_{{\mathcal R}} \biggl(\frac{k}{k_{p}}\biggr)^{n_{s} -1}, \qquad k_{p} = 0.002\, \mathrm{Mpc}^{-1},
\label{ST1Ca}
\end{equation}
where $k_{p}$ is called the pivot scale, $n_{s}$ is the scalar spectral index and ${\mathcal A}_{R}$ is the amplitude of the scalar power 
spectrum at the pivot scale.

The conformal rescaling of Eq. (\ref{EIN1}) assumes that the refractive index is homogeneous. If this is not the 
case the results for the tensor modes will be exactly the same while the scalar modes will be modified since 
the fluctuations of the conformal factor contribute to the scalar and not to the tensor problem. More specifically,
the scalar modes of the geometry in the Einstein frame: 
\begin{equation} 
\delta_{s} g^{(E)}_{00} = 2 a_{E}^2 \phi_{E},\qquad \delta_{s} g^{(E)}_{i j} 
= 2 a_{E}^2( \psi_{E} \delta_{ij} - \partial_{i}\partial_{j } C_{E}),\qquad \delta_{s} g^{(E)}_{0i} = - a_{E}^2 \partial_{i} B_{E},
\label{SSS1}
\end{equation}
will be related to the scalar modes in the conformally related frame via the scalar fluctuation of Eq. (\ref{G1}), i.e. 
\begin{equation}
\delta_{s} g_{\mu\nu}^{(E)} = \delta_{s} n \, \overline{G}_{\mu\nu}^{(s)} + n\, \delta_{s} G_{\mu\nu}.
\label{SSS2}
\end{equation}
Now the scalar fluctuations in the new frame can always be parametrized as 
in Eq. (\ref{SSS1}) but with different arbitrary functions: 
\begin{equation} 
\delta_{s} G_{00} = 2 a^2 \phi,\qquad \delta_{s} G_{i j} 
= 2 a^2 ( \psi \delta_{ij} - \partial_{i}\partial_{j } C),\qquad \delta_{s} G_{0i} = - a^2 \partial_{i} B.
\label{SSS3}
\end{equation}
Equation (\ref{SSS2}) implies a specific relation between the perturbed 
components in the two frames, i.e. 
\begin{equation}
\phi = \phi_{E} - \frac{1}{2} \biggl(\frac{\delta_{s} n}{n}\biggr),\qquad \psi = \psi_{E} + \frac{1}{2} \biggl(\frac{\delta_{s} n}{n} \biggr), \qquad C = C_{E}, \qquad B = B_{E}.
\label{SSS4}
\end{equation}
Note, however, that the curvature perturbations on comoving orthogonal hypersurfaces 
are frame-invariant. Indeed in the two frames they are simply given by 
\begin{equation}
{\mathcal R}_{E} = - \psi_{E} - \frac{{\mathcal H}_{E}}{n^{\prime}} \, \delta_{s} n ,\qquad 
{\mathcal R} = - \psi  - \frac{{\mathcal H}}{n'}\,\delta_{s} n.
\label{SSS5}
\end{equation}
Equation (\ref{SSS5}) does not imply that ${\mathcal R}_{E} \neq {\mathcal R}$, as it could be superficially concluded: 
since  $\psi_{E} \neq \psi$ (see Eq. (\ref{SSS4})) Eqs. (\ref{G5}) 
and (\ref{G7a}) imply  ${\mathcal H}_{E} \neq {\mathcal H}$. However, 
the mismatch between $\psi_{E}$ and $\psi$ is exactly compensated by the mismatch between 
${\mathcal H}_{E}$ and  ${\mathcal H}$. In fact,
from Eq. (\ref{G5}) we have $2 ({\mathcal H}_{E} - {\mathcal H})= n^{\prime}/n $ so that Eq. (\ref{SSS5}) implies 
${\mathcal R}= {\mathcal R}_{E}$. We therefore have, as anticipated, that the tensor modes of the geometry 
discussed in the bulk of the paper and the curvature perturbations on comoving orthogonal hypersurfaces are both 
frame-invariant and gauge-invariant. 

\end{appendix}


\newpage 

\begin{thebibliography}{99}

\bibitem{zeroa} L.~P.~Grishchuk,   Sov.\ Phys.\ JETP {\bf 40}, 409 (1975)   [Zh.\ Eksp.\ Teor.\ Fiz.\  {\bf 67}, 825 (1974)].

\bibitem{zerob}  L.~P.~Grishchuk, Annals N.\ Y.\ Acad.\ Sci.\  {\bf 302}, 439 (1977).

\bibitem{glauber} B.~R.~Mollow and R.~J.~Glauber,  Phys.\ Rev.\  {\bf 160}, 1076 (1967); Phys.\ Rev.\  {\bf 160}, 1097 (1967).

\bibitem{zeroc} L.~P.~Grishchuk, Class.\ Quant.\ Grav.\  {\bf 10}, 2449 (1993). 

\bibitem{zerod} A.~A.~Starobinsky, JETP Lett.\  {\bf 30}, 682 (1979) [Pisma Zh.\ Eksp.\ Teor.\ Fiz.\  {\bf 30}, 719 (1979)]. 

\bibitem{zeroe} V. A. Rubakov, M. V. Sazhin and A. V. Veryaskin, Phys. Lett. {\bf 115B}, 189 (1982).

\bibitem{zerof} M.~Giovannini, Mod.\ Phys.\ Lett.\ A {\bf 32}, no. 35, 1750191 (2017) [arXiv:1709.00914 [gr-qc]].

\bibitem{onea}  P.~Szekeres,  Annals Phys.\  {\bf 64}, 599 (1971).

\bibitem{oneb} P.~C.~Peters,  Phys.\ Rev.\ D {\bf 9}, 2207 (1974).

\bibitem{two} M.~Giovannini, Class.\ Quant.\ Grav.\  {\bf 33}, 125002 (2016)  [arXiv:1507.03456 [astro-ph.CO]].

\bibitem{three} Y.~Cai, Y.~T.~Wang and Y.~S.~Piao,
  Phys.\ Rev.\ D {\bf 93}, 063005 (2016) [arXiv:1510.08716 [astro-ph.CO]]; Phys.\ Rev.\ D {\bf 94},  043002 (2016)
 [arXiv:1602.05431 [astro-ph.CO]].

\bibitem{LIGO1} B.~P.~Abbott {\it et al.} [LIGO Scientific and Virgo Collaborations],  Phys.\ Rev.\ Lett.\  {\bf 119}, 141101 (2017) [arXiv:1709.09660 [gr-qc]].

\bibitem{LIGO2} 
  B.~P.~Abbott {\it et al.} [LIGO Scientific and Virgo Collaborations],  Phys.\ Rev.\ Lett.\  {\bf 119}, 161101 (2017) [arXiv:1710.05832 [gr-qc]].
  
 \bibitem{LIGO3} B.~P.~Abbott {\it et al.} [LIGO Scientific and VIRGO Collaborations],   Phys.\ Rev.\ Lett.\  {\bf 118}, 221101 (2017) [arXiv:1706.01812 [gr-qc]].

\bibitem{BICPL} P.~A.~R.~Ade {\it et al.} [BICEP2 and Keck Array Collaborations],
  Phys.\ Rev.\ Lett.\  {\bf 116}, 031302 (2016)  [arXiv:1510.09217 [astro-ph.CO]].

\bibitem{first}  M.~Giovannini,  Phys.\ Lett.\ B {\bf 759}, 528 (2016)  [arXiv:1603.09217 [astro-ph.CO]].

\bibitem{four}  L. H. Ford and L. Parker, Phys. Rev. D {\bf 16},1601 (1977).

\bibitem{four2}   L. H. Ford and L. Parker, Phys. Rev. D {\bf 16}, 245  (1977);  
B. L. Hu and L. Parker, Phys. Lett A {\bf 63}, 217 (1977).

\bibitem{dd1} W.~Zhao and Y.~Zhang, Phys.\ Rev.\  D {\bf 74}, 043503 (2006) [astro-ph/0604458].

\bibitem{dd2} Y.~Zhang, W.~Zhao, T.~Xia and Y.~Yuan, Phys.\ Rev.\  D {\bf 74}, 083006 (2006) [astro-ph/0508345].

\bibitem{absolute} M.~Giovannini, Class.\ Quant.\ Grav.\  {\bf 26}, 045004 (2009)  [arXiv:0807.4317 [astro-ph]];  
  Phys.\ Rev.\ D {\bf 82}, 083523 (2010) [arXiv:1008.1164 [astro-ph.CO]].

\bibitem{abr1} M. Abramowitz and I.A. Stegun, {\it Handbook of Mathematical Functions} (Dover, New York, 1972).

\bibitem{abr2} A. Erdelyi, W. Magnus, F. Obehettinger, and F. Tricomi, {\it Higher Trascendental Functions} (McGraw-Hill, New York, 1953).

\bibitem{FS1}   S.~Weinberg,  Phys.\ Rev.\  D {\bf 69}, 023503 (2004) [astro-ph/0306304].

\bibitem{FS2} D.~A.~Dicus and W.~W.~Repko, Phys.\ Rev.\ D {\bf 72}, 088302 (2005) [astro-ph/0509096].

\bibitem{violation} M.~Giovannini, Phys.\ Rev.\ D {\bf 89}, 123517 (2014) [arXiv:1404.7333 [hep-th]].

\bibitem{second} M.~S.~Turner, M.~J.~White and J.~E.~Lidsey,  Phys.\ Rev.\  D {\bf 48}, 4613 (1993).

\bibitem{absolute0} S.~Chongchitnan and G.~Efstathiou, Phys.\ Rev.\ D {\bf 73}, 083511 (2006) [astro-ph/0602594].

\bibitem{absolute2} M.~Giovannini,  Class.\ Quant.\ Grav.\  {\bf 31}, 225002 (2014) [arXiv:1405.6301 [astro-ph.CO]].

\bibitem{absolute3} M.~Giovannini,  Phys.\ Lett.\ B {\bf 759}, 528 (2016)  [arXiv:1603.09217 [astro-ph.CO]].

\bibitem{bbn1} V.~F.~Schwartzmann, JETP Lett.\ {\bf 9}, 184 (1969).
  
\bibitem{bbn2} M.~Giovannini, H.~Kurki-Suonio and E.~Sihvola, Phys.\ Rev.\  D {\bf 66}, 043504 (2002) [astro-ph/0203430].

\bibitem{bbn3} R.~H.~Cyburt, B.~D.~Fields, K.~A.~Olive, and E.~Skillman, Astropart.\ Phys.\ {\bf 23}, 313 (2005)  [astro-ph/0408033].

\bibitem{PUL1} V.~M.~Kaspi, J.~H.~Taylor, and M.~F.~Ryba,   Astrophys.\ J.\ {\bf 428}, 713 (1994).

\bibitem{PUL2} F.~A.~Jenet {\it et al.},  Astrophys.\ J.\  {\bf 653}, 1571 (2006) [astro-ph/0609013].

\bibitem{PUL3} W.~Zhao,  Phys.\ Rev.\ D {\bf 83}, 104021 (2011)  [arXiv:1103.3927 [astro-ph.CO]].

\bibitem{PUL4} P.~B.~Demorest {\it et al.},  Astrophys.\ J.\  {\bf 762}, 94 (2013) [arXiv:1201.6641 [astro-ph.CO]].

\bibitem{PUL5}  W.~Zhao, Y.~Zhang, X.~P.~You and Z.~H.~Zhu,  Phys.\ Rev.\ D {\bf 87},  124012 (2013)  [arXiv:1303.6718 [astro-ph.CO]].

\bibitem{PUL6}R.~M.~Shannon {\it et al.},  Science {\bf 349},  1522 (2015)  [arXiv:1509.07320 [astro-ph.CO]].

\bibitem{LV3} J.~Aasi {\it et al.} [LIGO Scientific and VIRGO Collaborations],  Phys.\ Rev.\ Lett.\  {\bf 113},  231101 (2014)
  [arXiv:1406.4556 [gr-qc]].
  
\bibitem{LV4} B.~P.~Abbott {\it et al.} [LIGO Scientific and Virgo Collaborations],  Phys.\ Rev.\ Lett.\  {\bf 118}, 121101 (2017)
  Erratum: [Phys.\ Rev.\ Lett.\  {\bf 119},  029901 (2017)]  [arXiv:1612.02029 [gr-qc]].
 
\bibitem{LV1} J.~Aasi {\it et al.} [LIGO Scientific Collaboration],  Class.\ Quant.\ Grav.\  {\bf 32}, 074001 (2015)  [arXiv:1411.4547 [gr-qc]].

\bibitem{LV2} F.~Acernese {\it et al.} [VIRGO Collaboration],  Class.\ Quant.\ Grav.\  {\bf 32}, 024001 (2015)  [arXiv:1408.3978 [gr-qc]].

\bibitem{LISA} P.~Amaro-Seoane {\it et al.},  GW Notes {\bf 6}, 4 (2013).

\bibitem{BBO}  G. M. Harry, P. Fritschel, D. A. Shaddock, W. Folkner, E. S. Phinney, Class.\ Quant.\ Grav.\  {\bf  23} 4887 (2006).

\bibitem{DECIGO1} S.~Kawamura {\it et al.}, J.\ Phys.\ Conf.\ Ser.\  {\bf 120}, 032004 (2008).

\bibitem{DECIGO2}  S.~Kawamura {\it et al.},  Class.\ Quant.\ Grav.\  {\bf 28}, 094011 (2011).

\bibitem{TAMA}  M. Ando et al., Phys. Rev. Lett. {\bf 86}, 3950 (2001). 

\bibitem{GEO1}   B.~Willke {\it et al.},ÊÊClass.\ Quant.\ Grav.\  {\bf 19}, 1377 (2002).

\bibitem{GEO2} H.~Grote [LIGO Scientific Collaboration], Class.\ Quant.\ Grav.\  {\bf 27}, 084003 (2010).

\bibitem{kagra1} Y.~Aso {\it et al.} [KAGRA Collaboration],  Phys.\ Rev.\ D {\bf 88}, no. 4, 043007 (2013).

\bibitem{kagra2} K.~Somiya [KAGRA Collaboration],  Class.\ Quant.\ Grav.\  {\bf 29}, 124007 (2012).

\bibitem{ET1} B.~Sathyaprakash {\it et al.},  Class.\ Quant.\ Grav.\  {\bf 29}, 124013 (2012)
  Erratum: [Class.\ Quant.\ Grav.\  {\bf 30}, 079501 (2013)]

\end{thebibliography}
\end{document}